\documentclass{article}

\usepackage[english]{babel}
\usepackage{xcolor}

\usepackage[letterpaper,top=2cm,bottom=2cm,left=3cm,right=3cm,marginparwidth=1.75cm]{geometry}

\usepackage[numbers,square]{natbib}

\usepackage{amsmath}
\usepackage{amssymb}
\usepackage{graphicx}
\usepackage{authblk}

\usepackage{booktabs}
\usepackage{float} %
\usepackage{soul}
\usepackage[colorlinks=true,allcolors=blue]{hyperref}

\title{A broadband, individually addressable 2D-3D photonic integrated circuit for trapped-ion qubit control}

\author{%
\begin{center}
\small
Yiyang Zhi\textsuperscript{*,1,4}, Daniel Klawson\textsuperscript{*,1,4}, Bingran You\textsuperscript{*,2,4}, Michael Bareian\textsuperscript{*,3,4}, Elijah Mossman\textsuperscript{*,3,4}\
Chun-Yuan Fan\textsuperscript{1}, Arkadev Roy\textsuperscript{1,4}, Ke Sun\textsuperscript{2,4}, Jason Lee\textsuperscript{2,4}, Sung Cheol Yoon\textsuperscript{2,4}, Qiming Wu\textsuperscript{2,4}, Lai Jiang\textsuperscript{2}, Wenjun Ke\textsuperscript{2}, Weiwei Wu\textsuperscript{2}, Sirui Tang\textsuperscript{1}, Zachary Wall\textsuperscript{3,4},
Jiaxiang Wang\textsuperscript{3,4}, Louis Paul Romero\textsuperscript{1}, Sam Vizvary\textsuperscript{3,4}, Steven Diaz\textsuperscript{3,4}, Eric R. Hudson\textsuperscript{3,4}, Wesley C. Campbell\textsuperscript{3,4}, Hartmut Haeffner\textsuperscript{2,4},\
and Ming C. Wu\textsuperscript{1,4}

\vspace{0.5em}

\begin{center}
\emph{\textsuperscript{1}Department of Electrical Engineering and Computer Sciences, University of California, Berkeley, Berkeley, California 94720, USA} \\
\emph{\textsuperscript{2}Department of Physics, University of California, Berkeley, Berkeley, California 94720, USA} \\
\emph{\textsuperscript{3}Department of Physics and Astronomy, University of California, Los Angeles, Los Angeles, CA 90095, USA} \\
\emph{\textsuperscript{4}Challenge institute for Quantum Computation, University of California, Berkeley, Berkeley, California 94720, USA}
\end{center}

\vspace{0.5em}

\emph{\textsuperscript{*}These authors contributed equally to this work} \\
\emph{Correspondence:}
\href{mailto:yiyang_zhi3@berkeley.edu}{yiyang\_zhi3@berkeley.edu};
\href{mailto:dklawson@berkeley.edu}{dklawson@berkeley.edu}

\end{center}%
}

\date{}

\begin{document}

\maketitle

\begin{abstract}

Trapped ions provide a high-fidelity platform for quantum information processing, yet delivery of multiple, distinct wavelengths across large networks of interaction zones remains a bottleneck.
Conventional free-space light delivery lacks scalability, while on-chip grating couplers suffer from narrow operational bandwidth that increases circuit footprint and optical interfacing complexity.
Here we show a broadband photonic integrated circuit capable of addressing individual ion locations.
The circuit combines a planar waveguide lens with a micromirror fabricated using two-photon polymerization at wafer scale. Optical characterization demonstrates three individual addressing spots from $\lambda$ = 405 -- 880\,nm with -25\,dB average intensity crosstalk at  $5\,\mu\mathrm{m}$ pitch.
We trap $^{40}\mathrm{Ca}^{+}$ above the mirror to demonstrate preferential repumping and utilize $^{138}\mathrm{Ba}^{+}$ ions to characterize optical crosstalk far away from the mirror. Charging of the dielectric and silicon structures currently limits a full experimental demonstration of the devices’ capabilities but could be mitigated in future designs. This monolithic photonic architecture could bring broadband addressing in an on-chip modality to trapped-ion technology. More broadly, combining additive manufacturing with established quantum device fabrication techniques could open new possibilities for implementing advanced
quantum architectures.

\end{abstract}

\section{Introduction}
 Trapped-ion quantum processors have demonstrated the highest gate fidelities and longest coherence times of any quantum-computing platform~\cite{harty2014high, ballance2016high, Ransford2026, oxford_record}.
 However, scaling beyond the present state of the art of $\sim 100$ physical qubits exposes an optical engineering challenge: universal control of a single ion species requires multiple, distinct laser wavelengths typically spanning the near-ultraviolet (NUV) to the near-infrared (NIR) range for photoionization, Doppler cooling, repumping, shelving, gate operation, and state readout.
 These wavelengths must be delivered with micron-level spatial precision to every qubit in the computational register~\cite{wineland1998experimentalIssues,kielpinski2002architecture}.
 The bench-top optical assemblies presently used for ion control are likely to scale unfavorably with qubit count in footprint, vibration sensitivity, and alignment complexity~\cite{niffenegger2020multiWavelength,kwon2024multiSite}.

Integrated photonics has emerged as a compelling solution to address this optical bottleneck.
In this approach, laser light is routed through nanophotonic waveguides patterned alongside the trap and emitted through on-chip diffractive grating couplers.
Monolithically integrated photonic ion traps have now been demonstrated with  $^{40}\mathrm{Ca}^+$, $^{88}\mathrm{Sr}^+$, and $^{171}\mathrm{Yb}^+$, enabling multi-wavelength control~\cite{niffenegger2020multiWavelength}, multi-ion quantum logic operations~\cite{mehta2020multiIonLogic}, multi-zone operations~\cite{mordini2025multizone, kwon2024multiSite}, polarization-gradient cooling~\cite{corsetti2026polarizationGradientCooling}, and on-chip fluorescence collection~\cite{Knollmann2026}.
Despite these advancements, current architectures rely on grating couplers as the final beam delivery element, introducing important optical design and chip layout constraints.

The tightly focused emission of a grating coupler relies on a wavelength-scale, rapidly varying spatial phase profile of the light.
Since the grating emission angle, propagation constant, and aperture phase are all wavelength dependent, there exists a strong tradeoff between simultaneously achieving high-efficiency focusing and operational bandwidth~\cite{Song2019, Xiao2013, Sapra2019, marchetti2017gratingCouplers}.
For example, inverse-designed double-layer gratings can push efficiency above 95\% and crosstalk below $-36$\,dB, but only at a single wavelength~\cite{Shirao2022}.
Therefore, multiple gratings are required for beam delivery to the ions. In a future large-scale quantum processor using a quantum charge-coupled device (QCCD) architecture ~\cite{kielpinski2002architecture, Pino2021}, the number of gratings required will scale with the number of incoming routing waveguides at each zone, whereas that of our $\mu$-optics system in principle does not. We estimate, under the assumption that each segmented zone contains enough grating couplers to provide full control of all qubits within that zone, that our architecture will have a smaller on-chip footprint for qubits number $N \gtrsim 3$ (Supplementary Information).

In this work, we introduce a hybrid, guided-wave and free-space nanophotonic architecture fabricated via wafer-scale photolithography and additive manufacturing.
Our integrated optical system is capable of delivering individually addressable optical beams from $\lambda = 405$\,nm to $\lambda = 880$\,nm at a height of $75~\mu\mathrm{m}$ above the surface, covering all of the key wavelengths required for the full control of $\mathrm{Ca}^{+}$ or $\mathrm{Ba}^{+}$ ions~\cite{Lindenfelser2017VisibleIRCa,Hucul2017BaSpectroscopy}.
We fabricate the two- and three-dimensional (2D-3D) photonic integrated circuit (PIC) on 6-inch silicon wafers in a process that monolithically integrates conventional waveguides and 3D-printed optics with a surface Paul trap.
Then, we trap $^{40}\mathrm{Ca}^{+}$ and $^{138}\mathrm{Ba}^{+}$ ions above the fabricated PIC-integrated ion trap. The calcium ions are used to demonstrate preferential repumping of $^{40}\mathrm{Ca}^{+}$ using on-chip delivery of 866\,nm light, whereas the barium ions are utilized as an optical probe to measure long-range light leakage from the integrated photonics.
Our approach combines multiple wavelengths and multi-qubit control onto a single optical system, paving the way for a scalable optical control layer of trapped-ion quantum hardware.

\section{Device architecture}

\begin{figure}[htbp]
\centering
\includegraphics[width=1\linewidth]{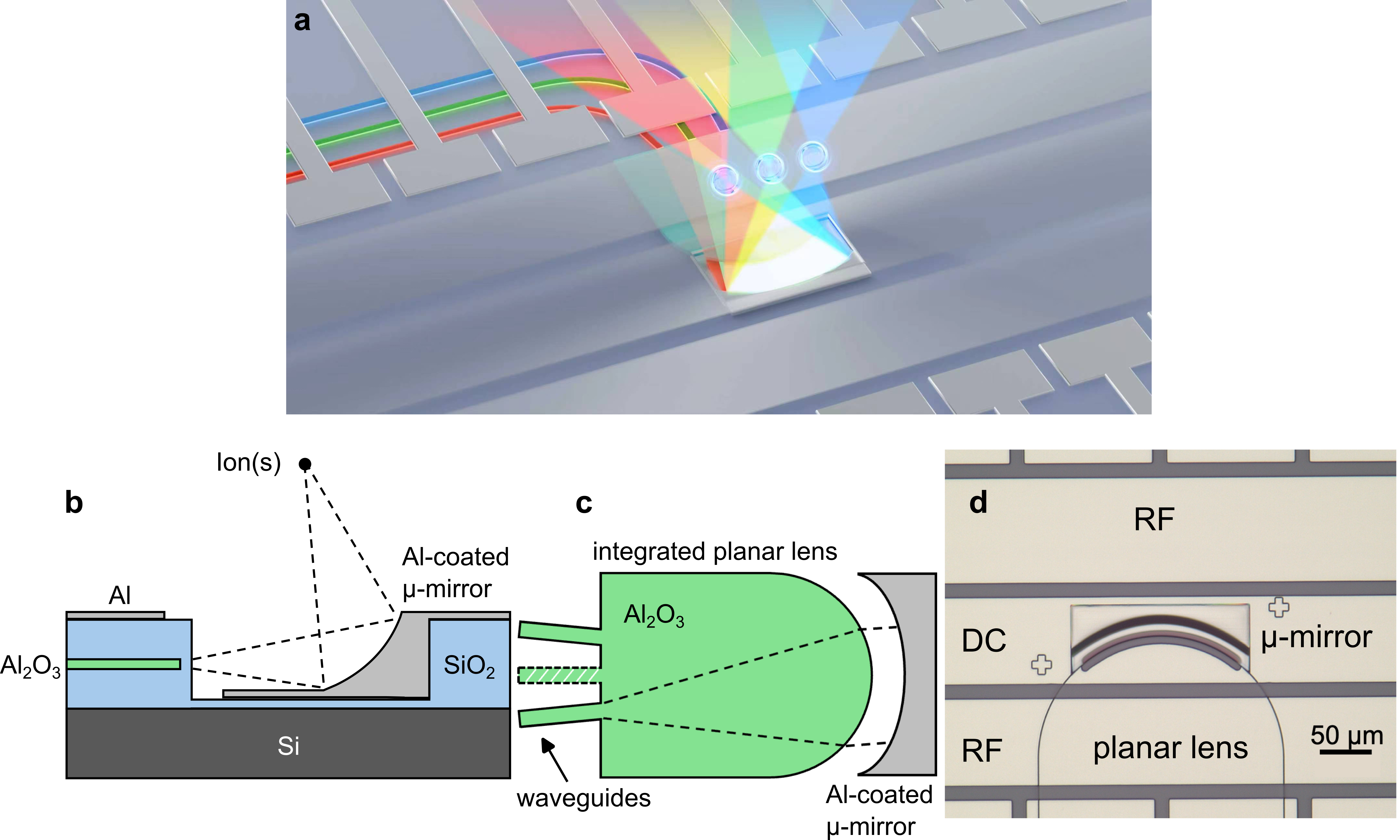}
\caption{\textbf{Architecture of the 2D-3D PIC.
a,} Isometric rendering of the device. The photonic layer is made up of an amorphous $\mathrm{Al_2O_3}$ core with a symmetric $\mathrm{SiO_2}$ cladding, patterned underneath the 200~nm Al surface-electrode layer. In the ion interaction region, a combination of lithographically defined photonics and 2PP-printed optics are used to route and shape the beam delivered to the ion location. \textbf{b,} Side view of the on-chip imaging system. The beam exits from the on-chip planar lens aperture and is focused by the Al-coated reflector to the ion location. \textbf{c,} Top view of the on-chip imaging system. The planar lens expands and collimates the optical mode from the incoming strip waveguides. The mirror focuses the optical energy onto the ion location. Each incoming waveguide is mapped to a different focal spot by the imaging system. The center dotted waveguide is not present in the calcium device design (Methods). \textbf{d,} Optical micrograph of a finished device showing the planar lens, the $\mu$-mirror, and the surrounding surface-trap electrodes. The mirror has the same potential as the center DC electrode. The alignment accuracy achieved between the substrate and the printed mirror is within $1~\mu\mathrm{m}$.}
\label{fig:concept_architecture}
\end{figure}

A rendering of our 2D-3D PIC is shown in Fig.~\ref{fig:concept_architecture}\textbf{a}.
The photonic stack consists of amorphous $\mathrm{Al_2O_3}$ (aluminum oxide) cladded symmetrically by $\mathrm{SiO_2}$ (silicon dioxide), which allows for broadband routing of light from NUV to NIR.
Previous works have demonstrated better than $3~\mathrm{dB/cm}$ propagation loss across 375--650\,nm in this material platform~\cite{west2019blueUV,mckay2023subDbAlumina}.
At the on-chip emission zone, guided light is first conditioned by a lithographically defined planar waveguide lens.
Upon exiting the waveguide, the beam is captured by a 3D biconic-aspherical micromirror ($\mu$-mirror) fabricated via two-photon polymerization (2PP), which redirects and focuses the light to a spot approximately 75~$\mu$m above the substrate surface.
A 5-wire surface trap~\cite{House2008} is fabricated directly atop the PIC using 200\,nm of Al (aluminum).
This metal layer serves a dual purpose: it acts as the reflective surface for the mirror and electrically connects it to the rest of the surface trap to confine the ion(s) at the focus of the optical beams.

The optical path of the system from the side view is illustrated in Fig.~\ref{fig:concept_architecture}\textbf{b}.
The light remains guided inside of the dielectric stack until the output of the waveguide lens, where it diverges with a numerical aperture ($\textrm{NA}\approx 0.3$) defined by the refractive index contrast between the guided mode and that of vacuum.
The $\mu$-mirror then redirects and focuses the narrow, rapidly diverging light onto the ion location.
In the in-plane direction (Fig.~\ref{fig:concept_architecture}\textbf{c}), the planar lens expands the optical mode from a sub-micron transverse extent in the strip waveguide to a $\approx 60~\mu\mathrm{m}$ aperture in the slab region, then collimates the mode via the circularly curved facet.
The $\mu$-mirror completes the imaging system by focusing the collimated output from the planar lens.
Each entrance waveguide maps to a discrete focal spot in the ion plane, axially spaced according to the pitch of the waveguides and the transverse magnification ($M_\text{T} \approx 0.5$) of the system. The center dotted waveguide is present for the barium but not the calcium device design (Methods).

An optical micrograph of a fabricated device is shown in Fig.~\ref{fig:concept_architecture}\textbf{d}.
The $\mu$-mirror, whose design is discussed in Methods, is printed flush within an etched trench at the center direct current (DC) electrode. This means the top surface of the mirror is at the same height as the surface electrodes and the sides of the mirror contact the trench walls. This reduces scattering if free-space beams by the 3D structure and avoids radio frequency (RF) power dissipation inside the polymer mirror.
The center DC electrode contains a metal cutout, clearing out the optical facet of the dielectric planar lens.
Alignment between the substrate and the printed optics is $\leq 1 \, \mu$m in all directions.
The potential of the mirror is the same as that of the center DC electrode to reduce distortion of the electric field produced by the $\mu$-mirror itself~\cite{Du2026}.

\section{Wafer-scale integration of 2D-3D optics with an ion trap}

\begin{figure}[htbp]
\centering
\includegraphics[width=1\linewidth]{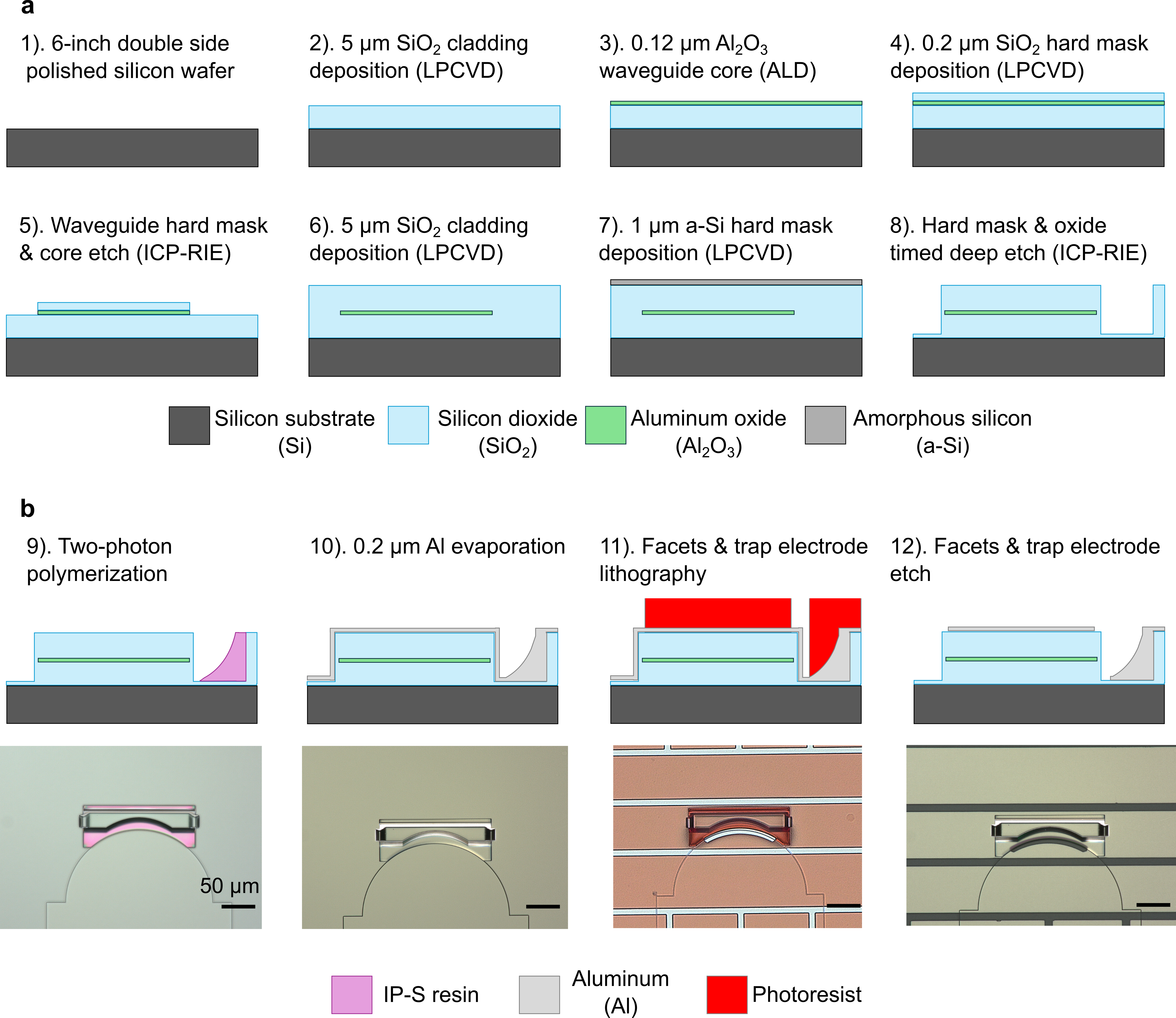}
\caption{\textbf{6-inch wafer-scale fabrication with integrated two-photon polymerized (2PP) optics.}
\textbf{a,} Pre-2PP process flow. The aluminum oxide waveguide core is grown via ALD for high optical quality. The silicon dioxide cladding is grown via LPCVD at $450^\circ\mathrm{C}$. The optical facets are defined through ICP-RIE, with amorphous silicon serving as the hard mask for the vertical sidewalls. \textbf{b,} Post-2PP process flow. We perform novel post-processing directly on the printed structures. It involves evaporation of aluminum for the surface electrodes and reflector, followed by contact lithography and aluminum etching to define the electrode gaps and optical facets. All scale bars on the microscope images are 50\,$\mu$m.  Finally, the devices are singulated via backside stealth dicing.}
\label{fig:fabrication}
\end{figure}

The full device is fabricated on a 6-inch silicon wafer platform.
An integrated photonics layer is defined in the pre-2PP process (Fig.~\ref{fig:fabrication}\textbf{a}). %
The process fabrication flow begins with $5~\mu\mathrm{m}$ of low-pressure chemical vapor deposition (LPCVD) of silicon dioxide, followed by 120\,nm of thermal atomic layer deposition (ALD) of aluminum oxide, patterned with a deep-UV stepper through an $\mathrm{SiO_2}$ hard mask and etched with a $\mathrm{BCl_3}/\mathrm{Ar}$ chemistry adapted from established practice~\cite{west2019blueUV}.
Waveguides are then symmetrically clad with $5~\mu\mathrm{m}$ of LPCVD silicon dioxide.
Optical facets, for both edge couplers and the planar lens, are defined by a vertical $\approx 9~\mu\mathrm{m}$ cladding etch through an amorphous-silicon (a-Si) hard mask.
We use a low-pressure $\mathrm{Ar}/\mathrm{CF_4}/\mathrm{CHF_3}$ inductively coupled plasma reactive ion etch (ICP-RIE) recipe to produce $\approx 89^\circ$ sidewalls that reduce refractive deflection of the exiting beam.
After waveguide fabrication, wafers are shipped to a 2PP foundry (Printoptix GmbH) for printing of the 3D optics.
The resulting $\mu$-mirrors show a measured surface roughness of $\approx 13$~nm RMS and part-per-thousand shape deviation from the design.
Wafers are then returned to the cleanroom for post-processing.

In the post-2PP process flow (Fig.~\ref{fig:fabrication}\textbf{b}), we integrate surface electrode traps directly atop the printed optics, taking into account the thermal, mechanical, and chemical compatibility of 2PP resin (IP-S).
We electron-beam evaporate 200\,nm of aluminum at a controlled wafer temperature ($\approx 25 \, ^\circ\mathrm{C}$) with a rotated, tilted geometry of the wafer chuck to coat the $\mu$-mirror conformally; the Al then serves both as the mirror reflector and surface-trap electrodes.
Trap electrodes and the planar-lens facet clear-out are patterned in a thick i-line resist (AZ MiR 900) with $6.5~\mu\mathrm{m}$ nominal thickness and $15.5~\mu\mathrm{m}$ pooled thickness in the facet trenches, to accommodate the large wafer topography.
Metal is then wet-etched in a heated phosphoric/nitric/acetic acid solution, with resist protecting the mirror surface.
The microscope images shown are for a slightly different design where the back surface of the $\mu$-mirror does not contact etched trench; there is no difference in terms of post-processing for the two designs.
To the best of our knowledge, this is the first demonstration of high-resolution lithography and etch on top of 2PP micro-optics.
This innovation directly enables the monolithic integration of 3D micro-optics with surface trap electrodes.

\section{Broadband, micron-level focusing}

\begin{figure}[htbp]
\centering
\includegraphics[width=1\linewidth]{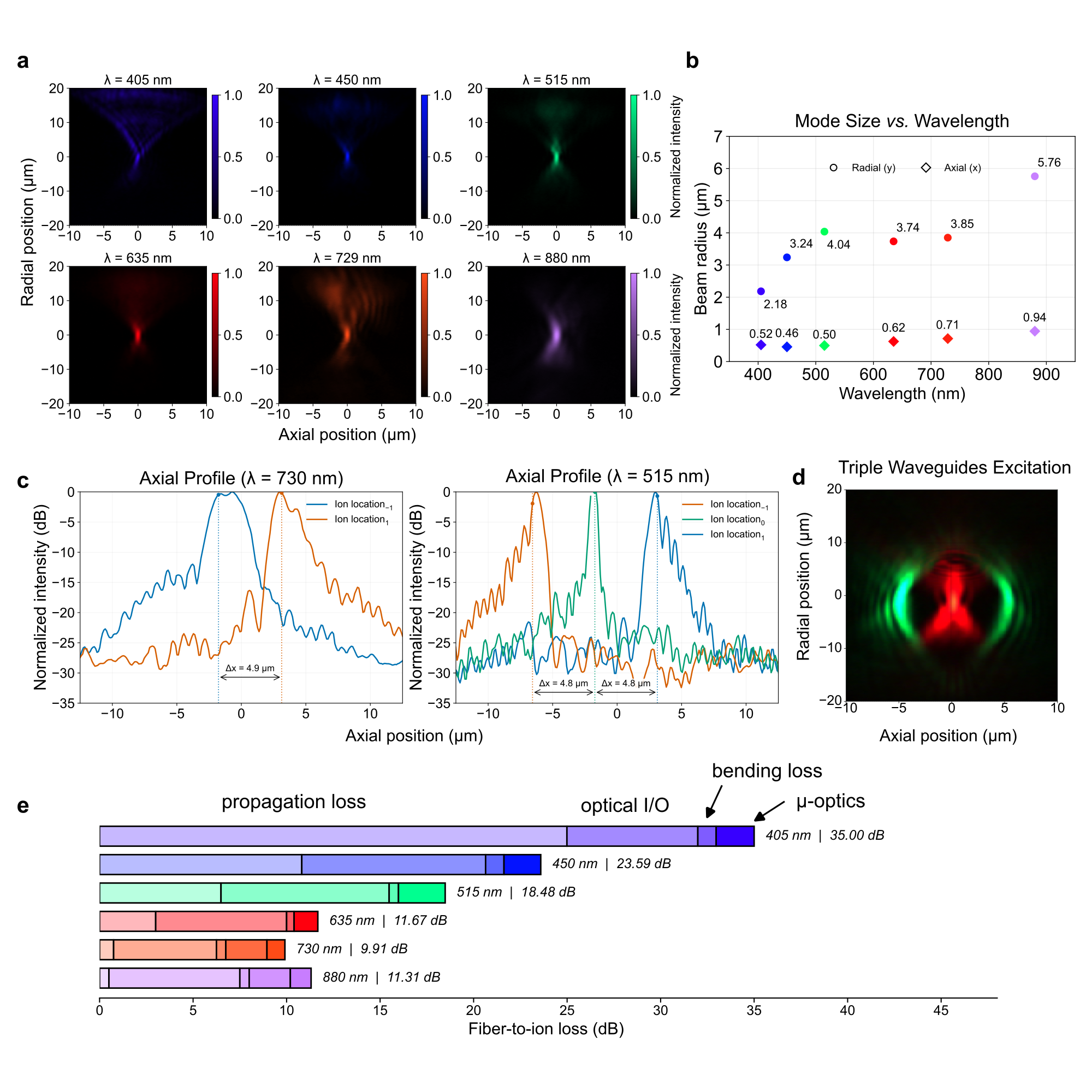}
\caption{\textbf{Broadband, micron-level focusing.}
\textbf{a,} Measured top-down focal-plane images at $h = 75\,\mu\mathrm{m}$ for six wavelengths from 405 to 880\,nm through the nanophotonic device. The shown color map for each wavelength is the perceived color at each wavelength except $\lambda=880$\, nm, which is a pseudo-color.
\textbf{b,} Axial (diamonds) and radial (circles) beam radius ($1/e^2$ intensity) versus wavelength. The effective NA is different in the axial and radial directions, and so is the achieved spot size.
\textbf{c,} Complete focal-plane intensity crosstalk for calcium device (two off-axis channels) and barium device (one on-axis channel and two off-axis channels). Excellent pitch ($5\,\mu\mathrm{m}$) registration is observed. The average any-to-any intensity crosstalk for the two wavelengths is -24.8\,dB.
\textbf{d,} Measured simultaneous three-spot illumination at 515\,nm and 635\,nm with a $\approx 5\,\mu\mathrm{m}$ pitch.
\textbf{e,} Measured fiber-to-ion loss across 405 nm -- 880\,nm, with bar segments from left to right showing the contributions of propagation loss, optical I/O loss (from the fiber/edge couplers interface), on-chip bending loss, and the insertion loss of the $\mu$-mirror itself. The color coding is the same as \textbf{a}.}
\label{fig:broadband_focusing}
\end{figure}

We characterize the optical emission using a piezo-actuated confocal profiler to re-image the emitted beam from our devices with input wavelengths spanning 405~nm to 880~nm (Methods).
For each wavelength, we acquire three-dimensional beam profiles by vertically translating the profiler in $1~\mu\mathrm{m}$ steps and capturing the intensity distribution at every height.
Fig.~\ref{fig:broadband_focusing}\textbf{a} shows the focal-plane intensity profile of a chip at $h = 75\,\mu\mathrm{m}$ -- the designed trapped-ion height -- for six measured wavelengths spanning from 405~nm to 880~nm.
The beam retains a tight, ``bowtie''-shaped focus across the entire band, demonstrating that a single optical element delivers all of the cooling, repump, shelving, and gate wavelengths needed for $\mathrm{Ca}^{+}$ and $\mathrm{Ba}^{+}$ operation.

Quantitatively, the device achieves micron-scale focusing along the trap axis, measured by the beam radius ($1/e^2$ intensity). The beam radius $w_x$ increases from $0.52~\mu\mathrm{m}$ at 405\,nm to $0.94~\mu\mathrm{m}$ at 880~nm (Fig.~\ref{fig:broadband_focusing}\textbf{b}).
The measured values are on average $12 \%$ higher than the predicted ones based on the image geometric $\mathrm{NA} \approx 0.34$, which we attribute to a non-uniformly illuminated aperture and spherical aberrations in our system.
Nevertheless, the axial focusing maintains its performance across approximately 500~nm bandwidth.
The radial beam radius $w_y$ remains below $6~\mu\mathrm{m}$ at all wavelengths and clusters near $4~\mu\mathrm{m}$, consistent with the asymmetric aperture set by the topography limit of the printed $\mu$-mirror.
Such asymmetry can be beneficial: in a linear ion trap, the ions typically are arranged along the axial direction with a qubit pitch of order 5\,$\mu$m. This geometry demands tight focusing along the axial direction for low crosstalk, whereas the radial direction can tolerate a broader beam. This larger radial beam improves robustness to ion-position uncertainty and to fabrication-induced beam-pointing errors.

To probe crosstalk relevant to multi-qubit control, we measured the any-to-any intensity crosstalk of the on-chip imaging system optically. For the calcium device, we have two off-axis entrance waveguides, whereas for the barium device, we have an additional on-axis waveguide. In the optical experiment, we excite individual entrance waveguides and monitor the intensity leakage at spatial locations corresponding to neighboring waveguides.
Figure~\ref{fig:broadband_focusing}\textbf{c} shows the measured focal-plane intensities at two representative wavelengths, 729\,nm for the optical qubit transition of $\mathrm{Ca}^{+}$, and 515\,nm that can be used to implement quantum gates with Raman transitions in $^{138}\mathrm{Ba}^{+}$.
In both cases, we observe that the imaging spots pitch agrees with the design to approximately the characterization imaging limit ($\approx 0.1 \mu$m per pixel). Moreover, the average intensity crosstalk falls below $-24$\,dB for any ion location pair. It differs slightly from the full-wave simulated crosstalk of $-30$\,dB potentially due to the metal roughness and stray light arising from the edge coupler facets. More detailed characterization on the optical crosstalk regarding channel-to-channel variation and tolerances can be found in Supplementary Information. To the best of our knowledge, this is the first demonstration of a device that can achieve individual addressing at two different wavelengths in a monolithic ion-trap PIC using a single shared optical element.

Our current architecture meets the requirement of addressing three ion locations along the trap, demonstrated experimentally with 515\,nm and 635\,nm excitation (Fig.~\ref{fig:broadband_focusing}\textbf{d}). The two green channels are driven coherently; however, due to their distinct spatial locations in the image plane, their profiles are not significantly influenced by each other despite interference patterns in the planar lens (Supplementary Information). The field of view is limited to $\approx 8^\circ$ by uncorrected coma. By introducing more optical elements, one can correct for these off-axis aberrations without sacrificing the per-emitter bandwidth gain.

We characterized end-to-end, fiber-to-ion optical loss with the chip packaged exactly as used in ion trapping experiments (Fig.~\ref{fig:broadband_focusing}\textbf{e}).
NIR losses are dominated by input-coupling and propagation, totaling $\sim 10$\,dB from fiber to focal spot.
Visible-band losses rise to 18.5~dB at 515~nm, and short-wavelength losses reach 24~dB at 450~nm and 35~dB at 405~nm.
Of these, the optical contribution of the planar-lens/mirror element itself is $\approx 2$\,dB across the entire band, extracted by subtracting independently measured propagation, bend, and coupling losses.
Short-wavelength performance is therefore not a limit of the architecture but of the alumina-on-silica waveguide platform, which can be substantially improved with established process refinements~\cite{mckay2023subDbAlumina}.

\section{Ion trapping results}

\begin{figure}[htbp]
\centering
\includegraphics[width=1\linewidth]{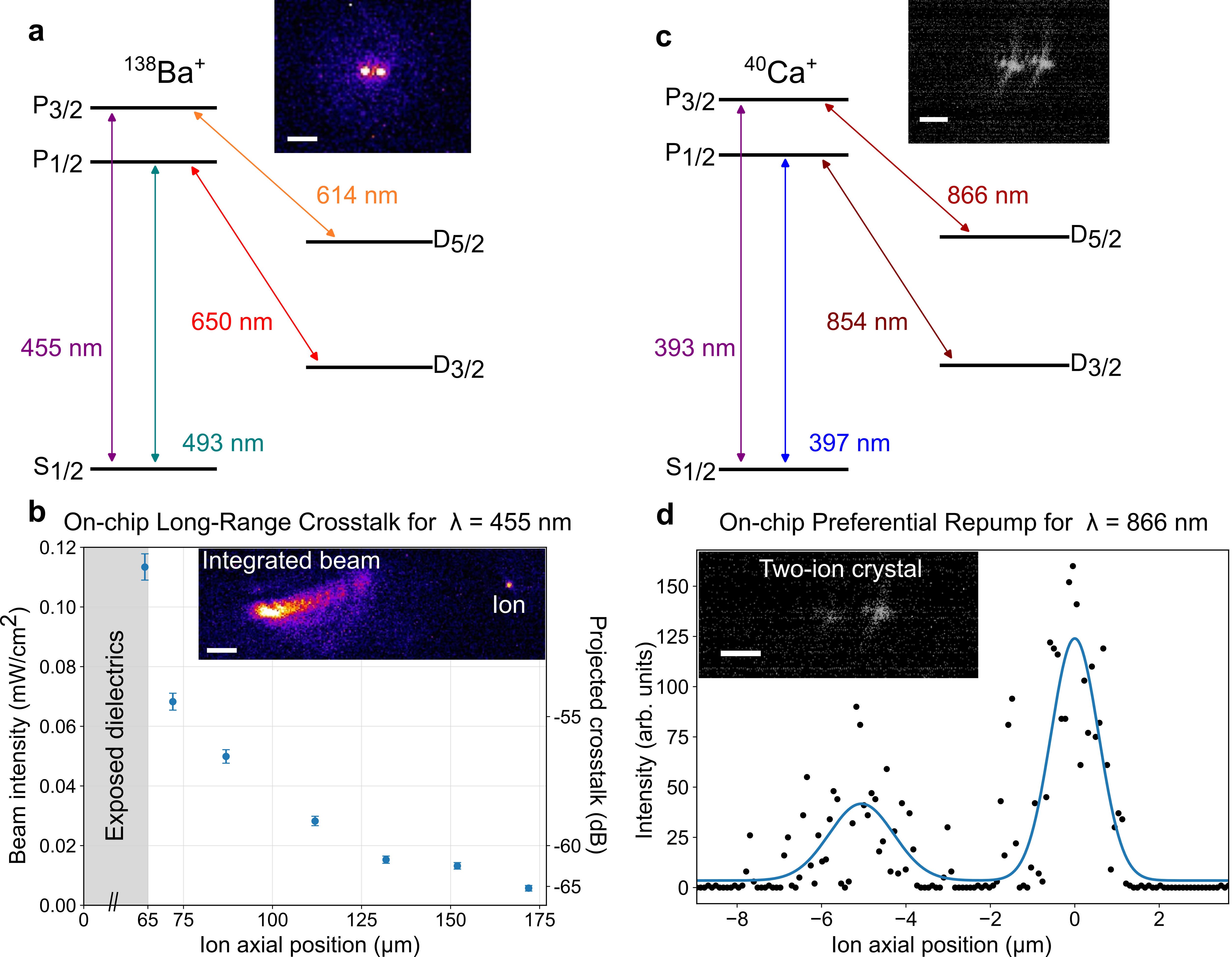}
\caption{\textbf{Ion trapping and crosstalk results.}
\textbf{a,} Relevant energy levels and optical transitions of
$^{138}\mathrm{Ba}^{+}$. The inset shows the fluorescence image of two trapped
$^{138}\mathrm{Ba}^{+}$ ions with free space 493\,nm and 650\,nm light shining onto the ions.
\textbf{b,} Measured 455~nm optical crosstalk intensity as a function of ion position near the dielectric cutout. Blue points show the intensity inferred from the measured excitation rate of the $S_{1/2}\rightarrow P_{3/2}$ transition of Ba$^+$, revealing that the intensity decreases with increasing distance from the dielectric mirror.
The shaded region indicates the dielectric/cutout region with its edge at $x=65~\mu\mathrm{m}$. The ion position is plotted relative to the center of the dielectric cutout ($x=0$). The inset shows a camera image of the ion and the integrated beam within the same field of view.
\textbf{c,} Relevant energy levels and optical transitions of
$^{40}\mathrm{Ca}^{+}$. The inset shows the fluorescence image of two trapped
$^{40}\mathrm{Ca}^{+}$ ions directly on top of the mirror using all free-space beams.
\textbf{d,} Demonstration of preferential repumping of one ion with integrated 866\,nm light delivered predominantly to one ion of a $5~\mu\mathrm{m}$-spaced Ca$^+$-ion pair. The plot shows the ion fluorescence intensity measured with a CCD-camera and two Gaussian fits showing the addressed ion fluorescence more strongly compared to its neighbor. The inset shows the camera image of two Ca$^+$ions above the mirror. All scale bars are 5\,$\mu$m.}
\label{fig:uhv_packaging}
\end{figure}

The integrated PIC was tested using both $^{138}\mathrm{Ba}^+$ and $^{40}\mathrm{Ca}^+$ ions.
Long-distance optical crosstalk was characterized using $^{138}\mathrm{Ba}^+$ ions, for which trapping directly above the $\mu$-mirror is still under active investigation, while illumination of $^{40}\mathrm{Ca}^+$ ions directly above the mirror through the integrated optics was demonstrated.

Trapping of $^{138}\mathrm{Ba}^{+}$ ions was demonstrated using the fabricated surface trap with free-space lasers (Fig.~\ref{fig:uhv_packaging}\textbf{a}).
Neutral barium atoms were generated by laser ablation of a sputtering target and photoionized (PI) using 413~nm and 531~nm light ~\cite{wall2025bariumautoionizationefficiention}.
The PI beams were triggered with the ablation pulse to reduce photoinduced charging, limiting exposure of the trap surface to less than 10~ms per loading attempt.
Strategies for shielding the exposed dielectric from photo-induced charging are discussed in the Discussion section.
Ions were loaded in the center of the trap and shuttled to different positions, with Doppler cooling and repumping performed via free-space 493~nm and 650~nm light, respectively.
The measured radial secular frequencies were repeatedly found to be 2.5 and 2.9~MHz,  corresponding to a RF voltage amplitude of $V_{\mathrm{rf}}=140~\mathrm{V}$, with ion lifetimes of one to two hours adjacent to the integrated $\mu$-mirror.

The crosstalk from the integrated 455~nm delivery path was characterized by translating the ion away from the beam center and measuring the rate of the $S_{1/2}\rightarrow P_{3/2}$ transition being driven, during which the ion is continuously laser cooled using the 493~nm cooling and 650~nm repump beams described in the previous paragraph.
The transition rate was determined from quantum-jump fluorescence traces by binning the bright-state dwell times and fitting their distribution to a Binomial process to extract the mean bright time, $\langle t_{\mathrm{bright}}\rangle$. The corresponding scattering rate, $(R_{\mathrm{sc}}=1/\langle t_{\mathrm{bright}}\rangle)$, was then converted to the local 455~nm intensity using the saturation parameter for the known detuning, beam waist, and input power.

During the measurement, $100~\mathrm{\mu W}$ of $455~\mathrm{nm}$ light, detuned by 200 MHz~from resonance, was coupled into the waveguide and output via the $\mu$-mirror.
A weak free-space 614~nm beam ($<$1~nW) was used to repump the population from the $D_{5/2}$ state. Figure~\ref{fig:uhv_packaging}\textbf{b} shows the measured 455~nm intensity as a function of the ion's axial position.
At a distance of $120~\mathrm{\mu m}$ from the expected peak center, the measured intensity is approximately $57~\mathrm{dB}$ below the expected peak intensity (inferred from offline measurement, see Fig.~\ref{fig:broadband_focusing}\textbf{e}) after accounting for the approximately $50\%$ population remaining in the $S_{1/2}$ state under strong saturation of the 493~nm transition.
This extinction is comparable to the weak-sidelobe levels characterized in precision grating-coupler optics, where relative intensities near $60~\mathrm{dB}$ were resolved tens of microns from the focus \cite{MehtaRam2017}.

Trapping of $^{40}\mathrm{Ca}^{+}$ above the $\mu$-mirror was achieved by photoionizing neutral calcium atoms emitted from a resistively heated oven using free-space 375~nm and 422~nm light (Fig.~\ref{fig:uhv_packaging}\textbf{c}).
The ions were Doppler cooled with free-space 397~nm Doppler cooling light while being shuttled axially toward the $\mu$-mirror with boundary element method (BEM)-derived DC voltages.
A stable two-ion crystal was achieved directly above the $\mu$-mirror with a lifetime of approximately one hour.

The integrated 866\,nm beam was shown to selectively excite a single ion of a two-ion crystal, with its neighbor 5~$\mu$m away showing weaker fluorescence (Fig.~\ref{fig:uhv_packaging}\textbf{d}).
This serves as a qualitative confirmation of the individual addressing capability measured from optical characterization. The ion fluorescence contrast is likely limited due to the combination of unoptimized axial positioning of the ion and saturation of the transition. In Supplementary Information, we show that axial misalignment on the order of $1\,\mu$m between the optical beam and the ion crystal coupled with the saturated response of the atom could render the fluorescence contrast to $\approx 5$ dB, our measured value. We also show that the intensity crosstalk remains robust to ion–beam misalignment of a few micrometers in the radial directions, suggesting that misalignment in these directions alone is unlikely to be the primary cause of the limited fluorescence contrast.

There are two principal challenges that we encountered while characterizing the fabricated devices: exposed dielectric charging in the barium experiment and trap instability in the calcium experiment. In the barium case, the exposed dielectric from the planar lens develops a positive voltage that weakens the axial confinement (see Supplementary Information), preventing us from shuttling the ion to the integrated beam. We believe that the $413$~nm and $531$~nm lasers could have contributed to the charging of the dielectric. In the calcium case, the micromotion compensation voltages in the out-of-plane direction dramatically increased throughout the chip during device characterization ($E_y: 2\,\text{V/mm} \rightarrow  6\,\text{V/mm}$ overnight after about two months). Subsequently, ions could not be loaded into the trap anymore to test performance at other wavelengths. This effect was observed throughout the chip, so it is unlikely connected to the $\mu$-mirror itself. It is hypothesized that a photo-induced change in conductivity of the silicon substrate could contribute to the loss of trap functionality~\cite{Mehta2014}. In the Discussion section, we consider methods to eliminate the effects of silicon charging by introducing a ground plane below the trap electrodes.

\section{Discussion}

In this work, we have demonstrated a hybrid 2D-3D photonic architecture for supporting the on-chip delivery of laser light to a chain of three trapped ion qubits.
Our system features broadband operation from $405$\,nm to $880$\,nm and supports individual addressing of qubit locations with an average any-to-any intensity crosstalk of about $-25$ dB at $5~\mu$m pitch.

We observe dielectric and semiconductor charging that limits trap performance, including poor motional coherence.
To reduce the effect of dielectric charging on trapping, a transparent, conducting oxide such as indium tin oxide (ITO) can be deposited on the planar lens facet.
Prior work has demonstrated stable trap operation~\cite{niffenegger2020multiWavelength, mordini2025multizone, Mehta2014} and high-fidelity two-qubit gates~\cite{mehta2020multiIonLogic} using ITO as a conducting layer.
Additionally, the 2PP platform can be leveraged to print a conductive shield -- a 3D-printed, metallized overhanging structure atop the planar lens that protects the ion from the fields generated by the exposed dielectric planar lens. From our simulation and calculation, the conductive shield could reduce the stray field coming from the planar lens by at least one order of magnitude.

To eliminate the effects of RF impedance fluctuations due to photo-generated carriers in silicon, a ground plane, either directly above the silicon handle layer~\cite{Mehta2014} or above the optically active dielectrics,~\cite{mehta2020multiIonLogic}, can be added to our material stack. The ground plane will terminate the RF field lines from reaching the silicon substrate.
It also shields charge carriers that are introduced in the silicon from influencing the ion qubits~\cite{Lee2024, Chung2025}.
Considering the fabrication thermal budget imposed by our photonics, we expect that with the addition of the ground plane above the waveguides routing layer, we will be able to operate the trap for an extended period of time.

Beyond the device-level improvements, we foresee favorable scaling of this architecture as the number of qubits in the system increases.
Since a single broadband mirror replaces many gratings and allows for individual addressing in a qubit register, the number of ions that can be controlled in a single zone with the same photonic footprint will increase, decreasing the number of ion transport needed per clock cycle in a QCCD machine~\cite{wineland2013nobelLecture,bruzewicz2019progress}.
Considering that the $\mu$-mirror is designed with full metallic coverage, this system provides a credible path to a densely populated trapped-ion processor in which the number of optical delivery elements only scales with the number of ion registers (not wavelengths) without imparting excessive charging and heating.
Taken together, the results presented here position hybrid 2D-3D photonics as a promising platform to further the scaling of ion trap quantum computers beyond $\sim 100$ qubits.

\section{Data availability}
The data that support the findings of this study are available from the corresponding authors upon reasonable request.

\section{Code availability}
The code used to analyze the measured data and model the saturated response the ion can be found in
https://doi.org/10.5281/zenodo.24289516.

\section{Acknowledgments}

This work was supported by the National Science Foundation through the Challenge Institute for Quantum Computation (CIQC) under grant number OMA-2016245. D.K. acknowledges support from the Army Research Office through the National Defense Science and Engineering Graduate (NDSEG) Fellowship Program. Y.Z. acknowledges support from the Hearts to Humanity Eternal (H2H8) Graduate Research Grant. M.B acknowledges support from the Army Research Office grant W911NF-20-1-0037 and National Science Foundation grants PHY-2207985 and PHY-2207546.

Wafer fabrication was carried out at the Marvell Nanofabrication Laboratory at the University of California, Berkeley, and the authors gratefully acknowledge the expert process and equipment staff for their support. The authors further acknowledge the Berkeley Sensor and Actuator Center (BSAC) and the Berkeley Emerging Technology Research Center (BETR).

The authors thank Printoptix GmbH for 2PP mirror process development, Optelligent LLC for fiber packaging engineering, SQS Vláknová optika for the vacuum-compatible fiber array units, Disco Hi-Tec America for stealth dice process development, and ficonTEC GmbH for fiber coupling support. We also thank Dr. Jianheng Luo, Dr. Johannes Henriksson, and Dr. Xiaoxing Xia for valuable discussions.

\section{Author contributions}

D.K. and C.F. conceived the photonics design. S.T. gave suggestions to the $\mu$-optics design. D.K. and Y.Z. fabricated the devices. Y.Z., A.R., and D.K. measured the beam profiles of the fabricated devices. Y.Z., D.K., and L.R. deployed the ion traps at UCB and UCLA. B.Y., K.S., J.L., S.Y., Q.W., L.J., W.K., and W.W. performed the quantum experiments for $^{40}\textrm{Ca}^+$ at UCB. M.B., E.M., Z.W., J.W., S.V., and S.D. performed the quantum experiments for $^{138}\textrm{Ba}^+$ at UCLA. E.H., W.C., H.H., and M.W. supervised the project. Y.Z., D.K., B.Y., M.B., E.M., and J.W. wrote the manuscript with input from all authors.

\section{Conflict of interests}

The authors declare no conflict of interests.

\section{Methods}
\subsection*{Routing photonics design}

Waveguide modes were computed in Ansys Lumerical Finite Difference Eigenmode (FDE) for the three operational bands. Single-mode operation across the 375--866\,nm window required band-specific widths at a common 120\,nm $\mathrm{Al_2O_3}$ thickness embedded in $5~\mu\mathrm{m}$ of $\mathrm{SiO_2}$ cladding. The three wavelength groups are near ultraviolet (NUV), visible (VIS) and near infrared (NIR). See Tab.~\ref{tab:target-wavelengths} for the wavelengths supported by each width of the optical waveguides. The thickness of the cladding is chosen such that the absorptive loss of the supported mode into the underlying silicon and overlying aluminum at all wavelengths is minimized.
\begin{table}[htbp]
    \caption{Target wavelengths for each spectral band.}
    \label{tab:target-wavelengths}
    \centering
    \begin{tabular}{lc}
        \hline
        Spectral band & Target wavelengths (nm) \\
        \hline
        NUV ($w = 0.55\,\mu\mathrm{m}$) & $375,\,397,\,422$ \\
        VIS ($w = 0.8\,\mu\mathrm{m}$)  & $450,\,532,\,650$ \\
        NIR ($w = 1.6\,\mu\mathrm{m}$)  & $729,\,854,\,866$ \\
        \hline
    \end{tabular}
\end{table}

Low-loss Euler bends were designed at each wavelength band such that the simulated bend losses fell below 0.01~dB per $90^\circ$ turn. This is done by first adiabatically tapering the routing waveguides to ones with a wider cross-section to increase the confinement factor, performing the bend, and tapering back to the propagation width. See Tab.~\ref{tab:bend-parameters} for the parameter of the designed Euler bands at each wavelength band.
\begin{table}[htbp]
    \centering
    \caption{Waveguide bend parameters for each wavelength band.}
    \label{tab:bend-parameters}
    \begin{tabular}{lccc}
        \hline
        Spectral band
        & \shortstack{Bend width\\($\mu\mathrm{m}$)}
        & \shortstack{Radius of curvature\\($\mu\mathrm{m}$)}
        & \shortstack{Euler parameter\\$p$} \\
        \hline
        NUV & 1 & 125 & 0.135 \\
        VIS & 2 & 175 & 0.135 \\
        NIR & 3 & 400 & 0.135 \\
        \hline
    \end{tabular}
\end{table}

The specific routing is slightly different for the calcium and barium device. For the former, two sets of NUV and NIR waveguides are each combined via an integrated wavelength division multiplexer (WDM) and fed into the common, integrated imaging system. Therefore, there are two off-axis entrance waveguides spaced such that the focusing spots they generate are $5\, \mu$m apart in the ion plane. For the latter, as the intended wavelengths fall within one wavelength band (VIS), three individual waveguides (two off-axis and one on-axis) are directly fed into the integrated imaging system without WDM. Integrated WDM between NUV and NIR were designed by adiabatic shaping of an asymmetric directional-coupler pair, with coupler length and gap selected to maintain better than 25~dB of suppression for the NUV mode while transferring NIR with sub-0.1~dB simulated loss.

\subsection*{Guided optics design}

At the high level, the planar lens is designed such that it produces a collimated beam in the in-plane directio and a diverging beam in the out-of-plane direction. The lens geometry was optimized in Lumerical 3D finite-difference time-domain (FDTD) with the 729\,nm wavelength used as the optimization target. In the future, multiple wavelengths could enter the optimization process for a composite planar lens to further reduce the residual chromatic aberration. Given computing resource constraints, the lens geometry was optimized at a small scale and verified by linearly scaling the optimized planar lens and fitting the collimation angle and mode field diameter as the optical system increased in size. See Tab.~\ref{tab:planar-lens-parameters} for the parameters of the planar lens geometry at the final, full scale.
\begin{table}[htbp]
    \centering
    \caption{Planar lens design parameters.}
    \label{tab:planar-lens-parameters}
    \begin{tabular}{lc}
        \hline
        Parameter & Value \\
        \hline
        Entrance waveguide width & $1.1\,\mu\mathrm{m}$ \\
        Slab width & $280\,\mu\mathrm{m}$ \\
        Slab length & $140\,\mu\mathrm{m}$ \\
        Lens radius of curvature & $98\,\mu\mathrm{m}$ \\
        Conic constant, $k$ & $0$ \\
        \hline
    \end{tabular}
\end{table}
The offset of the entrance waveguide relative to the center of the slab regions depends on the desired focal spots spacing ($5\, \mu$m in this case) and transverse magnification ratio ($M_{\text{T}}$) of the entire imaging system (planar lens and the micromirror, see below) as $\text{offset} = \frac{5 \, \mu\text{m}}{M_{\text{T}}}$. The angle of the entrance waveguide is selected such that it always points at the center of the planar lens to minimize the off-axis aberrations.

\subsection*{Micromirror design}
\label{sec:micromirror}
At the high level, the $\mu$-mirror takes the in-plane, collimated beam and out-of-plane, diverging beam produced by the planar lens and focus the light at the desired ion location. The biconic-aspherical $\mu$-mirror was optimized in Ansys Zemax. The planar-lens output mode was modeled in Zemax as a custom paraxial source with anisotropic divergence, $1^\circ$ lateral and $18^\circ$ vertical, and aperture, $75~\mu\mathrm{m} \times 2~\mu\mathrm{m}$, matched to the FDTD output. The reflector surface was parameterized as a biconic with independent $x$- and $y$-axis curvatures and conic constants, plus a 16-term aspherical term. Optimization proceeded in two stages: first, we only vary the first four biconic parameters to set the emission angle and ion height; then, we refine the remaining aspherical coefficients to maximize the modulation transfer function, weighted toward the sagittal direction (mapped to axial direction of the trap). We note that the spacing between the planar lens and the $\mu$-mirror is chosen such that the light in the vertical direction fills the mirror aperture to maximize the numerical aperture without clipping. The mirror has enough in-plane aperture such that there is negligible clipping for both on- and off-axis beams.

Optimized designs were re-imported into Lumerical FDTD and full waveguide-to-focal-plane simulations were performed at each operational wavelength (mentioned above) using UC Berkeley High Performance Computing (Savio) cluster.

\subsection*{Fabrication}

Devices were fabricated at the UC Berkeley Marvell Nanolab on 6-inch silicon wafers. $5~\mu\mathrm{m}$ of $\mathrm{SiO_2}$ was deposited by LPCVD at $450^\circ\mathrm{C}$, (Tystar Tytan II). 120~nm of $\mathrm{Al_2O_3}$ was grown by thermal ALD at $300^\circ\mathrm{C}$ with TMA/$\mathrm{H_2O}$ precursors (Cambridge Fiji F200). A 200~nm $\mathrm{SiO_2}$ hard mask was deposited by LPCVD, patterned through a UV photoresist stack (AR3GSF-600 BARC and Dow UV210GS-0.3) using a 248~nm KrF stepper (ASML PAS-5500/300B) and etched in $\mathrm{Ar}/\mathrm{CF_3}/\mathrm{CHF_4}$ chemistry (AMAT Centura eMxP+). The pattern was transferred into the alumina with a 1:2 $\mathrm{BCl_3}:\mathrm{Ar}$ reactive-ion etch (AMAT Centura DPS).

Waveguides were clad with $5~\mu\mathrm{m}$ of LPCVD $\mathrm{SiO_2}$. Optical facets were defined by depositing $1~\mu\mathrm{m}$ of amorphous silicon, LPCVD at $550^\circ\mathrm{C}$,(Tytan 3800) patterning with Dow UV210GS-0.6 on the ASML stepper, etching the hard mask in $\mathrm{HBr}/\mathrm{Cl_2}$, and transferring the pattern into the $10~\mu\mathrm{m}$ $\mathrm{SiO_2}$ stack with a tuned $\mathrm{Ar}/\mathrm{CF_4}/\mathrm{CHF_3}$ etch. The etch endpoint was set by reflectometry to leave $0.5$--$1.5~\mu\mathrm{m}$ of oxide above the silicon substrate to preserve electrical isolation of the trap electrodes.

Wafers were shipped to Printoptix GmbH for 2PP printing of 50 micromirrors per wafer using the Nanoscribe Quantum X and Nanoscribe IP-S resin, aligned to alumina-etched marks. After return, mirrors were metallized by tilted, rotated e-beam evaporation of 200 nm Al at controlled wafer temperature, below $200^\circ\mathrm{C}$ throughout. Trap electrodes and planar-lens facet clear-outs were patterned with spin-coated AZ MIR 900 at a $6.5~\mu\mathrm{m}$ nominal thickness and exposed with i-line contact lithography (Karl Suss MA6), then wet-etched in a $50^\circ\mathrm{C}$ bath of 80\% $\mathrm{H_3PO_4}$, 5\% $\mathrm{HNO_3}$, 5\% $\mathrm{CH_3COOH}$, and 10\% $\mathrm{H_2O}$. Wafers were diced via backside laser stealth dicing (Disco Hi-Tec America).

\subsection*{Optical characterization}

Beam profiles were measured with a custom-built piezo-actuated microscope: a $50\times, 0.6$ NA infinity-corrected objective (Mitutoyo plan apochromat), a visible-light tube lens (Thorlabs TTL200) and a color complementary metal-oxide-semiconductor (CMOS) camera (EO-1312C). The microscope was mounted on a piezoelectric stage (Newport ESP300) with $1~\mu\mathrm{m}$ step resolution. Six fiber-pigtailed diode lasers spanning 405, 450, 515, 635, 730, and 880 ~nm injected 3--30~mW of single-mode power into the input fiber array. High-dynamic-range maps of focal-plane intensity, Fig.~\ref{fig:broadband_focusing}\textbf{(c)}, were obtained by multi-exposure stitching of five frames per wavelength.

\subsection*{UHV packaging}

Fiber array units (FAUs) were aligned and bonded to the PIC by Optelligent LLC. The FAU and PIC were mounted on a shared silicon base bonded with Epo-Tek 353ND. Symmetric fused-silica support blocks were attached to the FAU sidewalls with UV-curable Optocast 3408 VM, then reinforced with low coefficient of thermal expansion thermal epoxy, Masterbond EP21TCHT-1, cured at $120^\circ\mathrm{C}$. Packages survived four sequential vacuum bakes to 50, 80, 100, and $125^\circ\mathrm{C}$ with negligible coupling loss.

Vacuum chambers at UC Berkeley and UCLA used 6-inch stainless-steel octagons with 2.75-inch CF AR-coated viewports and a laser-welded Vacom fiber feedthrough. Pumping employed a 20~L/s ion pump, a titanium sublimation pump, a non-evaporable getter, and a turbo pump that was disconnected after bakeout. Base pressures below $1 \times 10^{-10}$~Torr were maintained for the duration of ion experiments.
To test the device under the realistic conditions of a trapped-ion experiment, we packaged devices using ultra-high-vacuum (UHV) compatible fiber array units (FAUs) (Waveguide Technologies and SQS Vláknová optika). Achieving stable optical coupling through a $125^\circ\mathrm{C}$ bakeout is non-trivial; along with Optelligent LLC, we developed a symmetric-support pedestal package that mounts the FAU on a shared silicon base together with the PIC, bonded with a UV-curable epoxy and reinforced by a low coefficient of thermal expansion (CTE) thermal epoxy. The fiber feedthrough -- selected after comparative outgassing tests of multiple commercial parts -- is a laser-welded FC/APC flange (Vacom); this combination reached and held base pressures below $1 \times 10^{-10}$~Torr at room temperature in two independent vacuum chambers at UCB and UCLA.

\clearpage
\setcounter{section}{0}
\setcounter{subsection}{0}
\setcounter{figure}{0}
\setcounter{table}{0}
\setcounter{equation}{0}
\renewcommand{\thesection}{S\arabic{section}}
\renewcommand{\thefigure}{S\arabic{figure}}
\renewcommand{\thetable}{S\arabic{table}}
\renewcommand{\theequation}{S\arabic{equation}}
\renewcommand{\theHsection}{supp.\arabic{section}}
\renewcommand{\theHfigure}{supp.\arabic{figure}}
\renewcommand{\theHtable}{supp.\arabic{table}}
\renewcommand{\theHequation}{supp.\arabic{equation}}
\begin{center}
{\Large\bfseries Supplementary Information\par}
\medskip
A broadband, individually addressable 2D-3D photonic integrated circuit for trapped-ion qubit control
\end{center}

\section{Footprint comparison of grating couplers and shared micro-optics}

We compare the estimated optical footprint of a grating-based
architecture with that of a shared $\mu$-optics system. The comparison assumes a nominal ion height of $h = 75\,\mu\mathrm{m}$
and target Gaussian beam-waist radii of
$w_x = 10\,\mu\mathrm{m}$ and $w_y = 1\,\mu\mathrm{m}$, similar to what has been demonstrated in the main text. We choose common ion species used in quantum information science: calcium($\mathrm{Ca}^+$), barium($\mathrm{Ba}^+$), strontium($\mathrm{Sr}^+$), and ytterbium($\mathrm{Yb}^+$). The included wavelengths satisfy the basic control protocol of cooling, (re)pumping, (de)shelving, and coherent control. Within each wavelength group, it is assumed that a common grating coupler is used.

To estimate the grating opening area for a wavelength $\lambda$, we calculate the back-propagated beam radius at the emission plane as
\begin{equation}
    W_i(\lambda)
    =
    w_i
    \sqrt{
        1+
        \left(
            \frac{\lambda h}{\pi w_i^2}
        \right)^2
    },
    \qquad i \in \{x,y\}.
    \label{eq:grating-backpropagation}
\end{equation}
We further assume normal emission and define a representative rectangular aperture extending to two times the beam radius on either side of the beam center to capture all of the energy:
\begin{equation}
    A_{\mathrm{g}}(\lambda)
    =
    16 W_x(\lambda) W_y(\lambda).
    \label{eq:grating-aperture-area}
\end{equation}

Assuming one independently addressed grating output per
wavelength group per ion, the summed grating aperture area
for a zone containing $N$ ions is
\begin{equation}
    A_{\mathrm{G}}(N)
    =
    N A_{\mathrm{G,ion}},
    \qquad
    A_{\mathrm{G,ion}}
    =
    \sum_{j=1}^{3}
    A_{\mathrm{g}}(\overline{\lambda}_j),
    \label{eq:grating-zone-area}
\end{equation}
where $\overline{\lambda}_j$ is the arithmetic average wavelength in the $j$-th wavelength group. On the other hand, the $\mu$-optics system approximately occupies
\begin{equation}
    A_{\mu}
    =
    150\,\mu\mathrm{m}
    \times
    50\,\mu\mathrm{m}
    =
    7{,}500\,\mu\mathrm{m}^2
\end{equation}
of opened trench space on chip. This footprint is shared among the addressed ions and wavelength groups, provided that the register lies within the usable field of the imaging system (we estimate that for our architecture, $N\geq7$ is achievable).

The table summarizing the representative footprint comparison between the two architectures are presented in Tab.~\ref{tab:grouped-wavelength-footprint}. The crossover ion number $N_{\text{c}}$ represents the minimum number of ions in each zone such that the $\mu$-optics system has the lower total chip footprint. As can be seen from the table, when $N \geq 3$, our architecture has a clear advantage.
\begin{table}[htbp]
    \centering
    \caption{
        Estimated optical areas for grouped-wavelength
        addressing of common trapped-ion species.
        Each wavelength group is assigned one grating.
        Areas are rounded to the nearest $\mu\mathrm{m}^2$.
    }
    \label{tab:grouped-wavelength-footprint}
    \setlength{\tabcolsep}{5pt}
    \renewcommand{\arraystretch}{1.3}
    \resizebox{\linewidth}{!}{%
    \begin{tabular}{lcccc}
        \hline
        Ion
        & \shortstack{Wavelength groups\\(nm)}
        & \shortstack{Grating aperture\\area per ion\\
                      ($\mu\mathrm{m}^2$)}
        & \shortstack{Shared $\mu$-optics\\area per zone\\
                      ($\mu\mathrm{m}^2$)}
        & \shortstack{Crossover\\$N_{\text{c}}$} \\
        \hline
        $\mathrm{Ca}^{+}$
        & $\{397\},\ \{729\},\ \{854,866\}$
        & $4{,}498$
        & $7{,}500$
        & $2$ \\

        $\mathrm{Ba}^{+}$
        & $\{455,493\},\ \{532,585\},\ \{614,650\}$
        & $3{,}335$
        & $7{,}500$
        & $3$ \\

        $\mathrm{Sr}^{+}$
        & $\{422\},\ \{674\},\ \{1033,1092\}$
        & $5{,}274$
        & $7{,}500$
        & $2$ \\

        $\mathrm{Yb}^{+}$
        & $\{355,369.5\},\ \{638\},\ \{935\}$
        & $4{,}437$
        & $7{,}500$
        & $2$ \\
        \hline
    \end{tabular}%
    }
\end{table}

It is important to point out that this comparison only includes the scaling of the final optical delivery elements. Thus, it does not account for tapers, routing, inter-element spacing, and electrodes-induced geometric constraints in either cases. Moreover, it does not consider additional beam directions (standing waves and Raman beams, for instance) required by particular control protocols. Nevertheless, this analysis shows that our $\mu$-optics system potentially could accommodate more qubits per zone in a future QCCD processor when the zone area fixed by electrode size. Even for architecture choices that only require individual addressing on the optical gate beam, multi-species operations likely mean that each zone needs to support about six wavelengths groups. In this circumstance, our $\mu$-optics system again has the potential to support more qubits per unit optical footprint.

\section{Detailed crosstalk characterization}

We evaluate the uniformity and sensitivity of intensity addressing crosstalk for the calcium and barium devices through optical characterization. Uniformity refers to the differences in crosstalk between each channel, and sensitivity refers to how the crosstalk changes when there are radial and height offsets from the nominal ion positions. These offsets could arise from fabrication imperfection and misalignment.

\begin{figure}[!htb]
    \centering
    \includegraphics[width=1\linewidth]{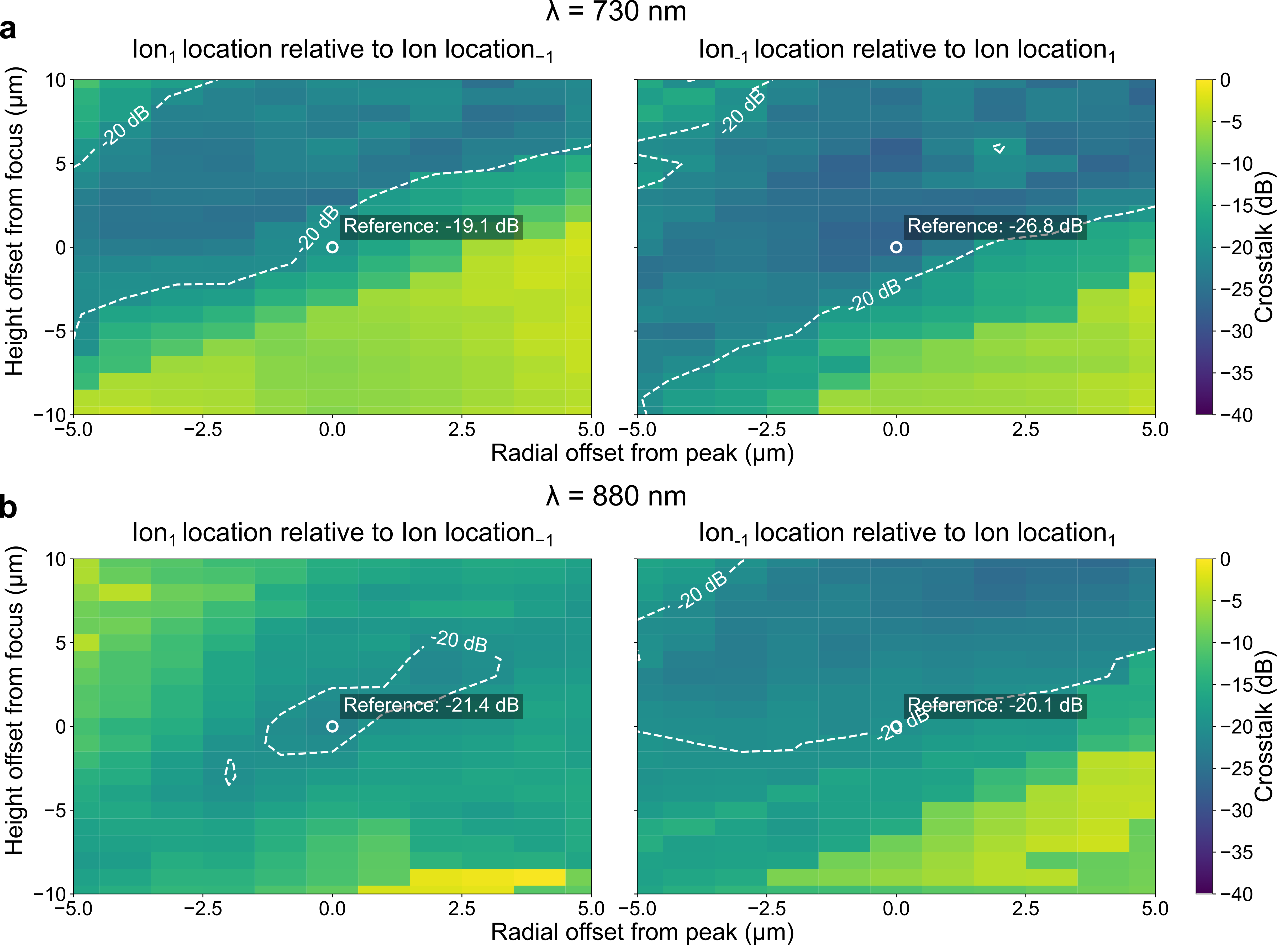}
    \caption{\textbf{Intensity crosstalk uniformity and tolerance for $\lambda =$ 729 (\textbf{a}) and 866 \text{nm} (\textbf{b})}. The data is taken at $\lambda =$ 730 and 880 \text{nm} as proxies for the atomci wavelengths. Overall, the channels are relatively consistent with each other on the $\pm 1 \,\mu$m level. The crosstalk is robust against radial shifts on the few microns levle.}
    \label{fig:ca_crosstalk}
\end{figure}

For the calcium device, we perform the crosstalk measurements for $\lambda = 730 \, \text{nm}$ and $\lambda = 880 \, \text{nm}$, which serve to project performances for $\lambda = 729, 854, 866 \, \text{nm}$ during actual qubit operations on the calcium ions. Fig.~\ref{fig:ca_crosstalk} presents the analysis plot for $730$\,nm (panel \textbf{a}) and $880$\,nm (panel \textbf{b}), respectively. For the optical qubit light, there exist deviations between the two off-axis channels, where one channel exhibits better crosstalk performance when the ion location is slightly better than $75 \, \mu$m. This behavior is consistent through multiple devices on the wafer, which we attribute to the small residual angle of the $\mu$-mirror relative to the planar lens during the printing process. Upon optical measurements of the improved devices with additional alignment marks, we observe that we can largely eliminate the angle misalignment effect. For the repump light, we still observe that one channel performs better than the other, evidenced by the area enclosed by the -20 dB contour lines. From our independent photonic measurements, our devices exhibit robustness to radial and height offset on the order of a few microns. Therefore, the $\approx 5$ dB suppression observed in our calcium experiment is likely a consequence of the saturated response of the ion and imperfect axial positioning of the qubit rather than the height (and radial) offset of the ion relative to the optical beam.

\begin{figure}[!htb]
    \centering
    \includegraphics[width=1\linewidth]{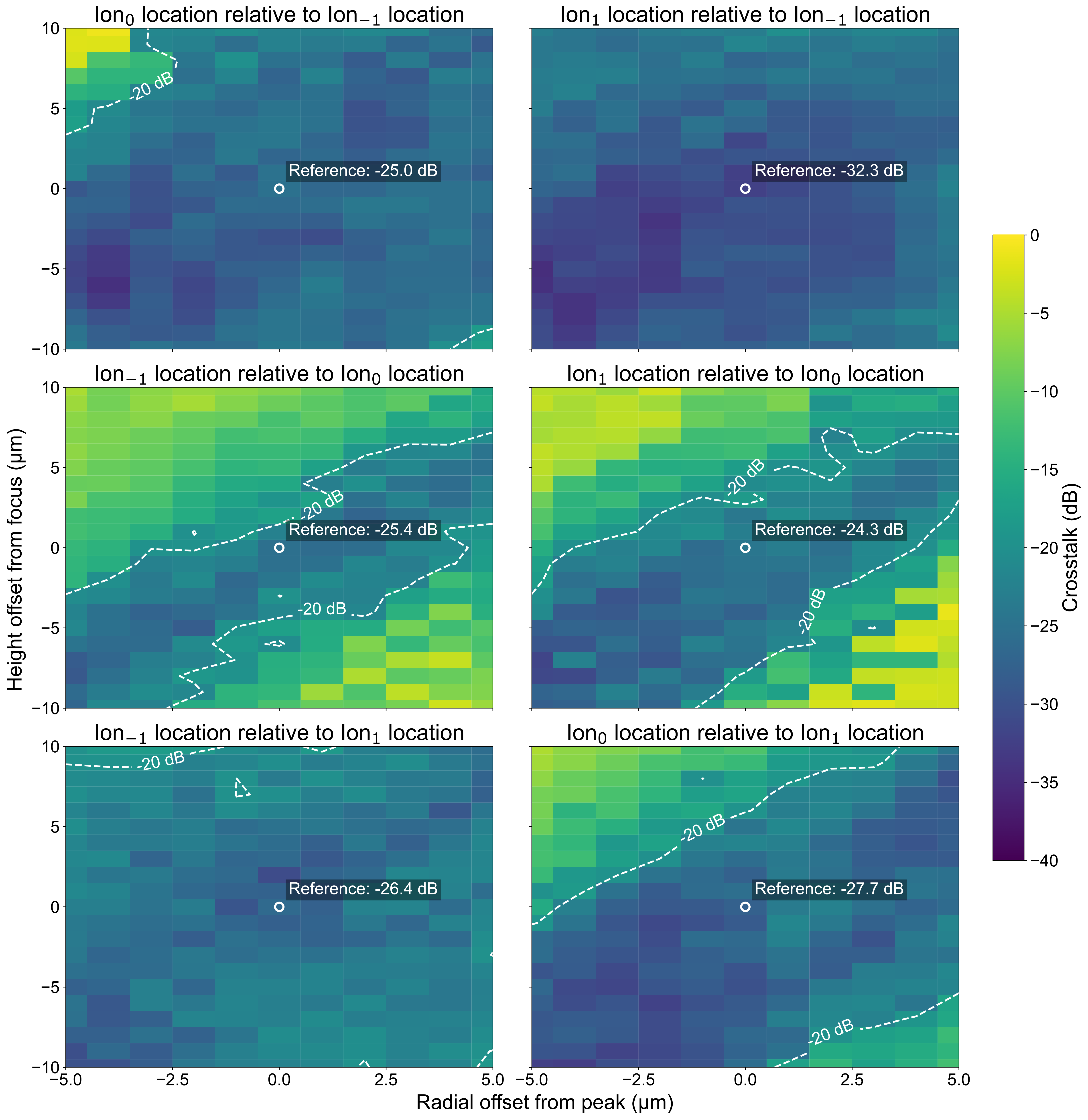}
    \caption{\textbf{Intensity crosstalk uniformity and tolerance for $\lambda = 515\, \text{nm}$}. All channels are relatively consistent with each other in in terms of crosstalk performance and are robust against misalignment errors in radial and height directions on the order of few microns,}
    \label{fig:ba_crosstalk}
\end{figure}

Fig.~\ref{fig:ba_crosstalk} presents the analysis plot for $\lambda = 515$\,nm, relevant for the barium ions. This wavelength can serve as the proxy for cooling, detection, repumping, and Raman gate light. The first row represents the intensity leakage to other ions locations from ion location$_{-1}$, second row ion location$_{0}$, and third row ion location$_{1}$. Similar to the calcium case, all addressing channels show consistent performance between the on-axis and off-axis beams. In terms of the performance tolerance, a range of a few microns is again observed. Furthermore, note that the -20 dB contour is larger for ion locations are physically further apart ($10 \,\mu$m as opposed to $5 \,\mu$m). Overall, our intensity crosstalk from optical characterization establishes our $\mu$-mirror platform as a strong candidate for integrated individual ion addressing.

\section{Effects of simultaneous input channels illumination}

We analyze the optical fields at an intended ion location when
multiple input channels are illuminated simultaneously. Because the photonic system operates in the linear regime, its response to simultaneous excitation is determined by the complex electric fields produced by the individual channels, guaranteed by Maxwell's equations.

Consider $N$ input channels operating at the same optical frequency.
Channel~1 is intended to address the ion at position $\mathbf{r}_0$,
whereas channels $2,\ldots,N$ contribute unwanted fields at that
position. Let $\mathbf{E}_j(\mathbf{r}_0)$ denote the complex electric
field produced by channel~$j$, including its excitation amplitude
and phase. The total field is
\begin{equation}
    \mathbf{E}_{\mathrm{tot}}(\mathbf{r}_0)
    =
    \sum_{j=1}^{N}\mathbf{E}_j(\mathbf{r}_0).
    \label{eq:total-field}
\end{equation}
We define the single-channel intensity leakage at the intended ion
location as
\begin{equation}
    \chi_j
    =
    \frac{I_j(\mathbf{r}_0)}{I_1(\mathbf{r}_0)}
    =
    \frac{
        \left\|\mathbf{E}_j(\mathbf{r}_0)\right\|^2
    }{
        \left\|\mathbf{E}_1(\mathbf{r}_0)\right\|^2
    },
    \qquad j=2,\ldots,N.
    \label{eq:single-channel-leakage}
\end{equation}
For an interaction sensitive to a particular dipole polarization, the corresponding projected electric fields should be used instead.

\subsection{General interference expression}

Expanding the squared total field gives
\begin{align}
    \frac{I_{\mathrm{tot}}}{I_1}
    &=
    1+\sum_{j=2}^{N}\chi_j
    \nonumber\\
    &\quad+
    2\operatorname{Re}
    \left[
        \sum_{j=2}^{N}
        \frac{
            \mathbf{E}_1^{*}\mathbf{E}_j
        }{
            \|\mathbf{E}_1\|^2
        }
    \right]
    \nonumber\\
    &\quad+
    2\operatorname{Re}
    \left[
        \sum_{2\leq j<k\leq N}
        \frac{
            \mathbf{E}_j^{*}\mathbf{E}_k
        }{
            \|\mathbf{E}_1\|^2
        }
    \right],
    \label{eq:vector-intensity}
\end{align}
where all fields and intensities are evaluated at $\mathbf{r}_0$.
The second term represents the sum of the individual intensity
leakages. The third term describes interference between the
intended field and the unwanted fields, whereas the fourth term
describes interference among the unwanted fields.

For fields with a common polarization, let $\phi_j$ denote the
phase of channel~$j$ relative to channel~1. Equation
\eqref{eq:vector-intensity} then becomes
\begin{align}
    \frac{I_{\mathrm{tot}}}{I_1}
    &=
    1+\sum_{j=2}^{N}\chi_j
    +
    2\sum_{j=2}^{N}
    \sqrt{\chi_j}\cos\phi_j
    \nonumber\\
    &\quad+
    2\sum_{2\leq j<k\leq N}
    \sqrt{\chi_j\chi_k}
    \cos(\phi_j-\phi_k).
    \label{eq:scalar-intensity}
\end{align}

\subsection{Mutually incoherent excitation}

If the channels are mutually incoherent over the relevant
measurement or interaction timescale, the interference terms
vanish upon averaging. The normalized intensity is then
\begin{equation}
    \frac{\langle I_{\mathrm{tot}}\rangle}{I_1}
    =
    1+\sum_{j=2}^{N}\chi_j.
\end{equation}
The fractional intensity increase at the intended ion location is
therefore
\begin{equation}
    \delta_{\mathrm{inc}}
    \equiv
    \frac{\langle I_{\mathrm{tot}}\rangle-I_1}{I_1}
    =
    \sum_{j=2}^{N}\chi_j.
    \label{eq:incoherent-increase}
\end{equation}
For approximately equal single-channel leakages,
$\chi_j\simeq\chi_0$, this reduces to
\begin{equation}
    \delta_{\mathrm{inc}}
    \simeq
    (N-1)\chi_0.
\end{equation}

\subsection{Coherent excitation and worst-case bound}

For coherent excitation, the intensity depends on the relative
phases and polarizations of the channels. Define
\begin{equation}
    \delta
    \equiv
    \frac{I_{\mathrm{tot}}-I_1}{I_1}.
\end{equation}
This quantity is the fractional intensity change at the addressed
ion and can be positive or negative. The worst-case bound is attained when all unwanted fields have the same polarization as the intended field and interfere constructively with it. Under these conditions,
\begin{equation}
    \delta_{\mathrm{max}}
    =
    \sum_{j=2}^{N}\chi_j
    +
    2\sum_{j=2}^{N}
    \sqrt{\chi_j}+
    2\sum_{2\leq j<k\leq N}
    \sqrt{\chi_j\chi_k}.
    \label{eq:constructive-maximum}
\end{equation}
If every unwanted channel satisfies $\chi_j\leq\chi_0$,
then
\begin{equation}
    \boxed{
    |\delta|
    \leq
    2(N-1)\sqrt{\chi_0}
    +(N-1)^2\chi_0.
    }
    \label{eq:equal-leakage-bound}
\end{equation}
Equality in the positive direction requires
$\chi_j=\chi_0$ for all unwanted channels, together with
constructive phase alignment and a common polarization.
When the total unwanted field amplitude is small,
\begin{equation}
    (N-1)\sqrt{\chi_0}\ll1,
\end{equation}
the quadratic contribution can be neglected, yielding the
leading-order worst-case expression
\begin{equation}
    \delta_{\mathrm{max}}
    \simeq
    2(N-1)\sqrt{\chi_0}.
    \label{eq:leading-coherent-increase}
\end{equation}
Thus, the leading coherent intensity perturbation scales with the
single-channel field leakage, $\sqrt{\chi_0}$, rather than the
single-channel intensity leakage, $\chi_0$. The actual perturbation can be smaller because of phase or
polarization differences.

\subsection{Numerical example}

For a single-channel intensity leakage of
\begin{equation}
    10\log_{10}\chi_0=-25~\mathrm{dB},
    \qquad
    \chi_0=10^{-2.5}\simeq3.162\times10^{-3},
\end{equation}
Tab.~\ref{tab:coherent-intensity-bound} lists the worst-case
fractional intensity increases calculated using the full
expression in Eq.~\eqref{eq:equal-leakage-bound}.

\begin{table}[htbp]
    \centering
    \caption{
        Worst-case fractional intensity increase at an addressed
        ion for simultaneous coherent excitation of $N$ channels,
        assuming $\chi_j=\chi_0=10^{-2.5}$ and fully
        constructive interference.
    }
    \label{tab:coherent-intensity-bound}
    \begin{tabular}{ccc}
        \toprule
        Number of channels, $N$
        & $\delta_{\mathrm{max}}$ (\%)
        & $10\log_{10}\delta_{\mathrm{max}}$ (dB) \\
        \midrule
        2 & 11.56 & $-9.37$ \\
        3 & 23.76 & $-6.24$ \\
        4 & 36.59 & $-4.37$ \\
        \bottomrule
    \end{tabular}
\end{table}

\section{Effect of axial positioning errors and repump saturation on fluorescence-based crosstalk measurements}

In the main text, we observe that the ion fluorescence measurement reveals a contrast on the order of 5 dB, while our photonic measurement suggests $\approx -25$ dB in intensity contrast. In this section, we consider the effect of axial positioning error and non-linear repump dynamics on the ion response contrast for a two-ion crystal spaced $\approx 5\, \mu$m apart.

\begin{figure}[H]
    \centering
    \includegraphics[width=1\linewidth]{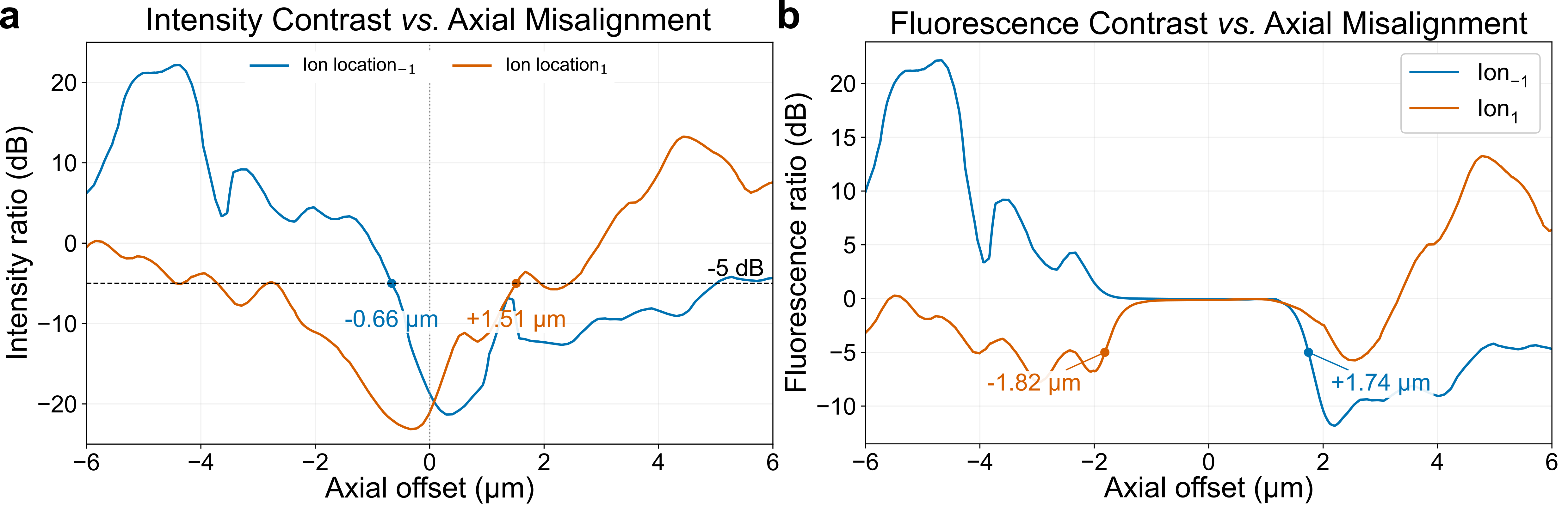}
    \caption{\textbf{Effects of axial misalignment}.
    \textbf{a,} Measured intensity contrast change of the beam $\lambda = 880$ nm as a function of axial offset from the ideal ion location placement from photonic characterization.
    \textbf{b,} Simulated ion fluorescence contrast change at $\lambda = 866$ nm as a function of axial offset from the ideal two-ion placement. In this case, the saturation parameter for the 397 nm light is on the order of 1. See text for the parameters and model used. }
    \label{fig:ca_response}
\end{figure}

Without considering the atom response, if the ion crystal is not placed ideally such that one ion is aligned with the addressing beam peak, the intensity contrast decrease itself may explain the fluorescence response of the ion. Fig.~\ref{fig:ca_response}\textbf{a} presents the analysis of $\lambda = 880$\,nm beam (proxy for $\lambda = 866$\,nm) when the two ion locations are placed away from the most optimized axial location (axial offset = 0\,$\mu$m). Since our beam waist is only on the order of a micron (Main), we observe that axial offset of the ion crystal on that order can lead to rapid decrease of intensity ratio, and this result is consistent regardless of which addressing channel is used.

As the next step, we include the response from the ion on top of the axial alignment to fully obtain the sensitivity of fluorescence contrast relative to axial misalignment. We consider continuous, simultaneous excitation of the $4S_{1/2}\leftrightarrow4P_{1/2}$ transition at
397~nm and repumping of the $3D_{3/2}\leftrightarrow4P_{1/2}$ transition at 866~nm in ${}^{40}\mathrm{Ca}^{+}$. The integrated photonic circuit delivers the 866~nm light, and the detected 397~nm fluorescence probes the repump response.

Fig.~\ref{fig:ca_response}\textbf{b} presents expected fluorescence contrast of the effective three-level model (described below) under axial offset using our experimental parameters in Tab.~\ref{tab:calcium-repump-parameters}.
\begin{table}[htbp]
    \centering
    \caption{
        Experimental parameters for the calcium repumping measurement. We assume $\pi$-polarized Zeeman coupling. Under these conditions, $\Tilde{s}_{866} \approx 159$ and $\Tilde{s}_{397} \approx 1$.
    }
    \label{tab:calcium-repump-parameters}
    \begin{tabular}{lcc}
        \hline
        Parameter & 866~nm repump & 397~nm excitation \\
        \hline
        Beam-waist radii
        & $1\times6~\mu\mathrm{m}$
        & $20\times100~\mu\mathrm{m}$ \\

        Input power to integrated photonics
        & $\approx50~\mu\mathrm{W}$
        & --- \\

        Integrated-path optical loss
        & $\approx12~\mathrm{dB}$
        & --- \\

        Optical power at the ions
        & $\approx3.15~\mu\mathrm{W}$
        & $\approx15~\mu\mathrm{W}$ \\

        Detuning magnitude, $|\Delta|/(2\pi)$
        & $0~\mathrm{MHz}$
        & $\approx20~\mathrm{MHz}$ \\
        \hline
    \end{tabular}
\end{table}
When the axial offset is smaller than the beam waist, the fluorescence contrast stays consistently at $\approx 0$\,dB since both ions in the crystal are strongly saturated. However, as soon as the axial misalignment reaches the scale of the beam radius, the florescence contrast ratios would increase, where we can recover the $\approx 5$\,dB contrast observed during our quantum experiment, consistent with our experiment and is independent of which addressing channel is used. In conclusion, through analysis of our photonic beam characteristics and ion experiment parameter, we determine that axial positioning errors on the order of the beam waist ($\approx 1 \, \mu$m) could be a factor to explain our apparent discrepancy between ion fluorescence contrast and photonic intensity contrast.

\subsection{Theoretical model}

\textbf{Effective three-level rate}. We denote the $4S_{1/2}$, $4P_{1/2}$, and $3D_{3/2}$ populations
by $p_S$, $p_P$, and $p_D$, respectively. Let $\Gamma$ be the total
spontaneous population decay rate of $4P_{1/2}$ and $b$ its
branching fraction into $3D_{3/2}$. We use $b=0.06435$ and
$1-b=0.93565$, consistent with the measured branching fractions
\cite{ca866_ramm2013}.

For an analytic estimate, we neglect coherent population trapping,
resolved Zeeman dynamics, and ion motion. We represent the two
driven transitions by effective bidirectional stimulated rates
$W_c$ and $W_r$. The population equations are
\begin{align}
    \dot p_S
    &=-W_c(p_S-p_P)+(1-b)\Gamma p_P,
    \label{eq:repump-rate-S}\\
    \dot p_D
    &=-W_r(p_D-p_P)+b\Gamma p_P,
    \label{eq:repump-rate-D}\\
    \dot p_P
    &=W_c(p_S-p_P)+W_r(p_D-p_P)-\Gamma p_P.
    \label{eq:repump-rate-P}
\end{align}
The rates $W_c$ and $W_r$ include stimulated absorption and
stimulated emission and are not the net photon-scattering rates.
In steady state, for nonzero $W_c$ and $W_r$,
\begin{equation}
    p_S=
    \left(1+\frac{(1-b)\Gamma}{W_c}\right)p_P,
    \qquad
    p_D=
    \left(1+\frac{b\Gamma}{W_r}\right)p_P.
\end{equation}
Using the normalization condition $p_S+p_P+p_D=1$ gives
\begin{equation}
    p_P=
    \left[
        3+\frac{(1-b)\Gamma}{W_c}
        +\frac{b\Gamma}{W_r}
    \right]^{-1}.
    \label{eq:repump-steady-population}
\end{equation}
The detected fluorescence rate is
\begin{equation}
    F=(1-b)\Gamma p_P,
    \label{eq:repump-fluorescence-rate}
\end{equation}
At fixed 397~nm excitation, define
\begin{equation}
    A=3+\frac{(1-b)\Gamma}{W_c},
    \qquad
    u=\frac{A W_r}{b\Gamma},
    \qquad
    F_\infty=\frac{(1-b)\Gamma}{A}.
\end{equation}
For fixed repump detuning and polarization,
$W_r\propto I_{866}$, so the effective parameter $u$ can be
written as
\begin{equation}
    u=\frac{I_{866}}{I_{1/2}},
\end{equation}
where $I_{1/2}$ is the effective repump half-saturation intensity
for the combined 397/866~nm excitation cycle. Consequently,
\begin{equation}
    F(I_{866})
    =
    F_\infty\frac{u}{1+u}
    =
    F_\infty\frac{I_{866}}{I_{866}+I_{1/2}}.
    \label{eq:repump-saturation-response}
\end{equation}
Here, $F_\infty$ is the fluorescence plateau of the reduced rate
model.

\textbf{Relation between intensity and fluorescence contrast}. Let
\begin{equation}
    c=\frac{I_{866,\mathrm{n}}}{I_{866,\mathrm{t}}}
\end{equation}
be the local 866~nm intensity ratio between the neighboring ion
and the addressed ion, denoted by subscripts $\mathrm{n}$ and
$\mathrm{t}$, respectively. We assume that the two ions
experience equal 397~nm excitation, detunings, and
polarization-dependent response, and that their relative
detection efficiencies have been corrected. Their effective
repump parameters then satisfy
\begin{equation}
    u_{\mathrm{n}}=cu_{\mathrm{t}}.
\end{equation}
From Eq.~\eqref{eq:repump-saturation-response}, the fluorescence
ratio is
\begin{equation}
    r_F
    \equiv
    \frac{F_{\mathrm{n}}}{F_{\mathrm{t}}}
    =
    \frac{c(1+u_{\mathrm{t}})}
         {1+cu_{\mathrm{t}}}.
    \label{eq:repump-fluorescence-ratio}
\end{equation}
For $0<c<1$ and $u_{\mathrm{t}}\geq0$,
\begin{equation}
    r_F-c
    =
    \frac{c(1-c)u_{\mathrm{t}}}
         {1+cu_{\mathrm{t}}}
    \geq0,
\end{equation}
and therefore
\begin{equation}
    c\leq r_F\leq1.
    \label{eq:repump-intensity-bound}
\end{equation}
Thus, under these assumptions, the fluorescence ratio provides
an upper bound on the intensity leakage. In the weak-repump
limit $u_{\mathrm{t}}\ll1$, fluorescence is approximately linear
in repump intensity and $r_F\simeq c$. In contrast, when both
$u_{\mathrm{t}}\gg1$ and $cu_{\mathrm{t}}\gg1$, both ions
approach the same fluorescence plateau and $r_F\rightarrow1$.

\textbf{Relation to transition saturation parameters}. The effective repump parameter $u$ is distinct from a bare
saturation parameter of the 866~nm transition. To specify the
convention explicitly, define
\begin{equation}
    s_j=\frac{2|\Omega_j|^2}{\Gamma^2},
    \qquad
    \widetilde{s}_j
    =
    \frac{s_j}{1+(2\Delta_j/\Gamma)^2},
    \qquad
    j\in\{397,866\},
\end{equation}
where $\Omega_j$ is the effective Rabi angular frequency,
including the relevant dipole coupling, and $\Delta_j$ is the
laser detuning in angular-frequency units.

With the additional approximation that the optical-coherence
decay rate is $\Gamma/2$, the stimulated transition rates are
\begin{equation}
    W_j=\frac{\Gamma}{2}\widetilde{s}_j.
\end{equation}
Substituting into the definition of $u$ gives
\begin{equation}
    u=
    \frac{
        3+2(1-b)/\widetilde{s}_{397}
    }{2b}\,
    \widetilde{s}_{866}.
    \label{eq:repump-bare-saturation}
\end{equation}

 \section{Static Fields from Charged Dielectrics}

A trapped ion provides an in situ probe of local static electric fields. We write the applied electrostatic potential as a linear superposition of the electrode basis potentials,

\[\Phi_{\mathrm{DAC}}(\mathbf{r})=\sum_j V_j \phi_j(\mathbf{r}),\] where $V_j$ is the voltage applied to electrode $j$, and $\phi_j(\mathbf{r})$ is the potential generated by that electrode for unit voltage, with all other electrodes grounded. At the micromotion null, the total DC electric field at the ion position must vanish. Therefore, the applied compensation field is equal and opposite to the local stray field such that \[
\mathbf E_{\mathrm{stray}}(\mathbf r)
+
\mathbf E_{\mathrm{comp}}(\mathbf r)
=
0.
\]
The stray field inferred from the compensation voltages is therefore \[
\mathbf E_{\mathrm{stray}}(\mathbf r)
=
-\mathbf E_{\mathrm{comp}}(\mathbf r)
=
\sum_j V_j \nabla \phi_j(\mathbf r).
\] The corresponding voltages needed to create arbitrary potentials are evaluated using BEM simulations.
Thus, the BEM-calculated electrode matrix, together with the experimentally determined compensation voltages, give the stray electric field at the ion position. By repeating this measurement as the ion is translated axially, we reconstruct the spatial dependence of the stray electric field near the dielectric boundary.

Micromotion compensation is performed by applying a small modulation to the RF drive near the secular frequencies of the ion and minimizing the driven response \cite{nadlinger}. Because excess micromotion is proportional to the ion displacement from the RF null, the modulation-induced response provides a sensitive measure of the residual DC field.

For each axial position, compensation voltages are adjusted until the response is minimized along the measured modes. These voltages are then converted into electric fields at the ion position using BEM.

Equivalently, the BEM solution provides the local multipole expansion of the applied potential about the ion position,
\begin{equation}
\begin{aligned}
    \Phi(x,y,z) &= C+ E_x(-x) + E_y(-y) + E_z(-z) \\
    &+ U_1(\frac{x^2 - y^2}{2}) + U_2 (2z^2 - x^2 - y^2) \\
    &+ U_3(\frac{xy}{2}) + U_4(\frac{yz}{2}) + U_5(\frac{xz}{2}) + O(r^3)
    \label{eq}
\end{aligned}
\end{equation}

The linear coefficients ($E_i$) determine the compensation field, while the quadratic terms determine the contribution of the applied DC voltages to the secular curvatures. In particular, $U_2$ determines the contribution to the axial curvature.

\begin{figure}
    \centering
    \includegraphics[width=1\linewidth]{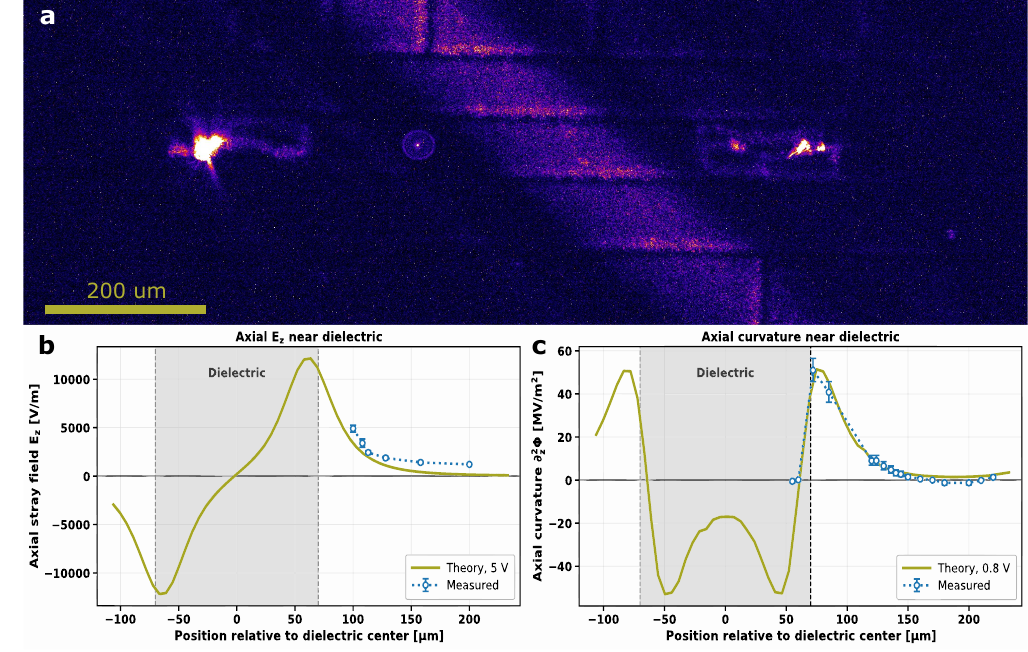}
    \caption{\textbf{Axial electrostatic perturbation from a charged dielectric}.
    \textbf{a,} Fluorescence image near the dielectric region.
    \textbf{b,} Measured axial stray field, $E_z$, compared with a BEM simulation.
    \textbf{c,} Axial potential curvature, $\partial_z^2\Phi$, extracted from secular-frequency measurements using $\partial_z^2\Phi=m(2\pi f_z)^2/q$.
    Blue points are experimental data with linewidth-derived uncertainties; gold curves are simulations modeling the charged dielectric surfaces as equal-voltage electrodes.
    The shaded region marks the dielectric.}
    \label{fig:charged-dielectric-curvature}
\end{figure}

Figure~\ref{fig:charged-dielectric-curvature}\textbf{b} shows the measured axial stray field as the ion is shuttled toward the dielectric boundary. The shaded region denotes the dielectric side of the cutout. The measured stray field corresponds to a positive charge on the dielectric, which produces a stray field of up to several thousand $\mathrm{V/m}$ within tens of microns of the dielectric boundary. This strong spatial dependence is consistent with a localized charged dielectric patch. The uncertainty in the reconstructed stray field is most likely dominated by the ion-position uncertainty. We propagate this uncertainty through the BEM field-response matrix using Monte Carlo simulation and take the standard deviation of the resulting field distribution as the error bar.

The charged dielectric also modifies the curvature of the trapping potential. A localized surface charge distribution can act as a ``pseudo-electrode''~\cite{House2008}, producing higher-order electrostatic terms in addition to a uniform stray field. These terms are observed through measurements of the ion secular frequencies, which are experimentally taken using parametric excitation of the RF. For a principal mode $i$, the curvature of the total trapping potential is related to the secular frequency by

\begin{equation}
    \begin{aligned}
        m\omega_i^2 &=q\frac{\partial^2 U_{\mathrm{tot}}}{\partial u_i^2}\\
        U_{tot} &= q\Phi_{DC} + U_{RF}
        \label{eq1}
    \end{aligned}
\end{equation}

where $m$ and $q$ are the ion mass and charge, $u_i$ is the coordinate along the principal axis of mode $i$, and the secular frequency comes from both the DC potential contribution and the RF pseudopotential. Equivalently, changes in the measured secular frequency provide a direct measure of changes in the local curvature.

Figure~\ref{fig:charged-dielectric-curvature}\textbf{c} shows the axial secular frequency measured over the same spatial range as the field data. The axial frequency changes by more than a factor of three near the dielectric boundary, indicating a factor of nine difference in the axial curvature. Changes in axial curvature have been shown to lead to heating in shuttling~\cite{burton2022multispecies}.

Modeling the exposed dielectric as an effective pseudo-electrode produces a large anti-confining axial curvature near the dielectric boundary, corresponding to an axial frequency shift on the order of several megahertz. The data in Figure~\ref{fig:charged-dielectric-curvature} corresponds to a value of 0.8 V on the dielectric. Since the geometric coupling between this electrode directly underneath the ion and the ion position is large, even a modest dielectric charge density can produce a non-negligible modification of the trapping potential. Simulations further indicate that the electrode voltages required to compensate for the resulting stray field can exceed typical operating limits near the dielectric, making it difficult to maintain a stable trapping potential while transporting the ion over or near the exposed region.

These measurements show that dielectric charging makes global compensation of stray fields difficult. Compensation at one axial position can leave residual field and curvature errors elsewhere, and large local stray fields can also rotate the principal axes of the trap, reducing Doppler-cooling efficiency. For scalable reflective photonics integrated with surface-electrode ion traps, these effects represent a significant technical challenge. Practical architectures, therefore, require both active charge control and electrostatic shielding of exposed dielectric materials.
\section{Long-range optical crosstalk measured from barium ions}
In an integrated QCCD processor, an operation beam addressed to one zone must not drive a transition in a neighboring zone. For photons at the relevant frequency of the trapped ion, even weak sidelobes can lead to optical pumping, fluorescence background, or coherent rotations of spectator ions.

Using the measured 455 nm crosstalk, we estimate the scattering-limited coherence time of an ion located near the dielectric cutout. The optical signal was recorded with a Thorlabs CS895MU camera using an exposure time of $15~\mathrm{ms}$ per frame.  The 455\,nm measurement can characterize the spatial falloff of stray light from the integrated photonics output: the intensity rises rapidly as the ion approaches the optical output region and remains low over the remote spectator range. To estimate the effect of residual 493~nm light on a spectator ion,  we assume that the same relative crosstalk profile applies at 493~nm and rescale the measured profile by the 493~nm intensity. For 493~nm light detuned from resonance of the $^{138}$Ba$^{+}$ $S_{1/2}\rightarrow P_{1/2}$ cooling transition by $\Delta/2\pi = 10~\mathrm{MHz}$, the inferred coherence time is approximately $100~\mathrm{\mu s}$ for an ion positioned about $120~\mathrm{\mu m}$ from the dielectric cutout based on:
\begin{equation}
\begin{aligned}
\Gamma_{\mathrm{sc}}(d)
&=
\frac{\Gamma}{2}
\frac{s(d)}
{1+s(d)+\left(2\Delta/\Gamma\right)^2},
\quad
T_{\mathrm{coh}}(d)
\approx
\Gamma_{\mathrm{sc}}^{-1}(d),
\end{aligned}
\end{equation}
where $\Gamma$ is the natural linewidth of the $P_{1/2}$ excited state, $s(d)$ is the saturation parameter defined as the real-time intensity divided by the saturation intensity at ion distance $d$, $T_{\mathrm{coh}}(d)$ is the coherence time, and $T_{\mathrm{sc}}(d)$ is the scattering rate. Though this demonstrates suppression of the integrated emission, additional suppression would be required to preserve coherence in spectator ions during gate operations.

\clearpage
\bibliographystyle{unsrtnat}
\bibliography{references}
\end{document}